\pdfoutput=1

\documentclass[11pt,twoside,a4paper,cmspaper,final,collab]{cms-tdr}
\def\svnVersion{215791}\def\cmsCernNo{2013-060}\def\cmsCernDate{\today}\def\cmsMessage{Published in the Journal of High Energy Physics as \href{http://dx.doi.org/10.1007/JHEP10(2013)062}{\doi{10.1007/JHEP10(2013)062.}}}

\begin{document}\cmsNoteHeader{FSQ-12-019}

\hyphenation{had-ron-i-za-tion}
\hyphenation{cal-or-i-me-ter}
\hyphenation{de-vices}

\RCS$Revision: 204993 $
\RCS$HeadURL: svn+ssh://svn.cern.ch/reps/tdr2/papers/FSQ-12-019/trunk/FSQ-12-019.tex $
\RCS$Id: FSQ-12-019.tex 204993 2013-08-30 12:45:34Z paoloa $
\providecommand{\pj}{\ensuremath{\mathrm{j}}\xspace}
\newlength\cmsFigWidth
\ifthenelse{\boolean{cms@external}}{\setlength\cmsFigWidth{0.85\columnwidth}}{\setlength\cmsFigWidth{0.4\textwidth}}
\ifthenelse{\boolean{cms@external}}{\providecommand{\cmsLeft}{top}}{\providecommand{\cmsLeft}{left}}
\ifthenelse{\boolean{cms@external}}{\providecommand{\cmsRight}{bottom}}{\providecommand{\cmsRight}{right}}
\cmsNoteHeader{FSQ-12-019} 
\title{Measurement of the hadronic activity in events with a Z and two jets and extraction of the cross section for the electroweak production of a Z with two jets  in pp collisions at $\sqrt{s} = 7\TeV$}

\date{\today}

\abstract{
The first measurement of the electroweak production cross section of a Z boson with two
jets (Zjj) in pp collisions at $\sqrt{s} = 7$\TeV is presented, based on a data sample
recorded by the CMS experiment at the LHC with an integrated luminosity of 5\fbinv.
The cross section is measured
for the  $\ell\ell \pj\pj $ ($\ell = \Pe, \Pgm$) final state
in the kinematic region
$m_{\ell\ell} >50$\GeV, $m_{\pj\pj} >120\GeV$,
transverse momenta $\pt^\pj  > 25$\GeV and
pseudorapidity $\abs{\eta^\pj}< 4.0$.
The measurement, combining the muon and electron channels, yields
                      $\sigma = 154
                      \pm 24\stat
                      \pm 46\,\text{(exp. syst.)}
                      \pm 27\,\text{(th. syst.)}
                      \pm 3\lum\unit{fb}$,
in agreement with the theoretical cross section.
The hadronic activity, in the rapidity interval between the jets,
is also measured. These results establish an important foundation
for the more general study of vector boson fusion processes, of relevance
for Higgs boson searches and for measurements of electroweak
gauge couplings and vector boson scattering.
}

\hypersetup{%
pdfauthor={CMS Collaboration},%
pdftitle={Measurement of the hadronic activity in events with a Z and two jets and extraction of the cross section for the electroweak production of a Z with two jets  in pp collisions at sqrt(s) = 7 TeV},%
pdfsubject={CMS},%
pdfkeywords={CMS, physics, Z boson}}

\hyphenation{cal-or-i-me-ter}

\maketitle 

\section{Introduction \label{sec:introduction}}
\begin{figure*}[htb]
\begin{center}
\includegraphics[width=0.30\textwidth]{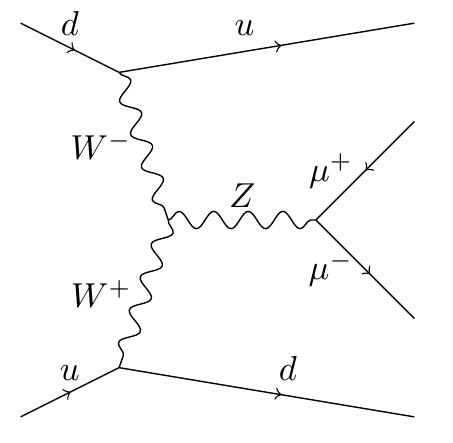}
\includegraphics[width=0.30\textwidth]{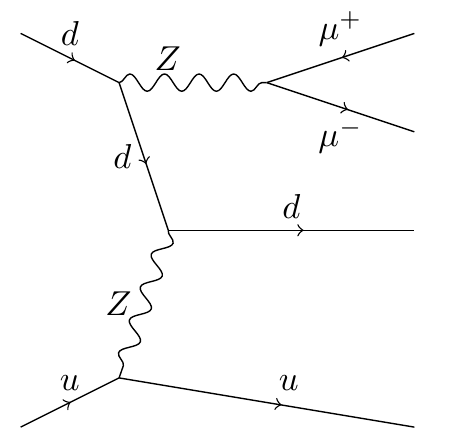}
\includegraphics[width=0.30\textwidth]{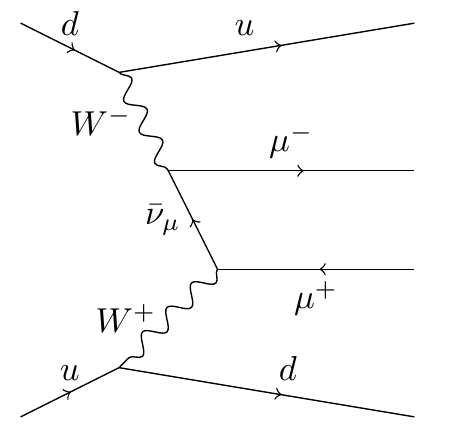}
\caption{Representative diagrams for EW $\ell\ell\Pq\Pq'$ production (for $\ell$$=$$\Pgm$):
VBF (left), bremsstrahlung (middle), and multiperipheral (right).}
\label{fig:introduction_EW}
\end{center}
\end{figure*}

\begin{figure*}[htb]
\begin{center}
\includegraphics[width=0.30\textwidth]{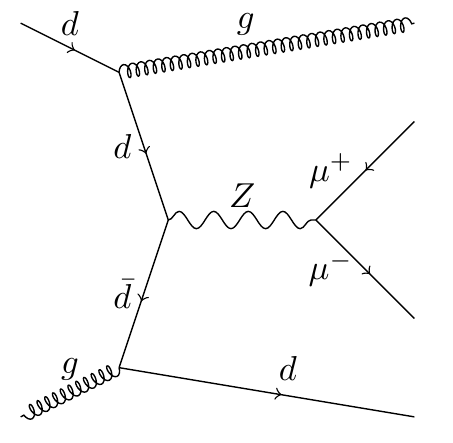}
\caption{Representative diagram for Drell--Yan production in association with two jets.}
\label{fig:introduction_DY}
\end{center}
\end{figure*}

The cross section for the electroweak (EW) production of a central W or Z boson in association
with two jets that are well separated in rapidity is quite sizable at the Large Hadron Collider
(LHC)~\cite{Oleari:2003tc}. These electroweak processes have been studied in the context
of rapidity intervals in hadron
collisions~\cite{Rainwater:1996ud,Khoze:2002fa}, as a probe of anomalous triple-gauge-boson
couplings~\cite{Baur:1993fv}, and as a background to Higgs boson searches in  Vector Boson Fusion
(VBF) processes~\cite{Rainwater:1998kj,Plehn:1999xi,Rainwater:1999sd,Kauer:2000hi}.
There are three classes of diagrams to be considered in the EW
production of $\PW$ and $\cPZ$ bosons with two
jets: VBF processes,  bremsstrahlung, and multiperipheral processes.
A full calculation reveals a large negative interference between the pure VBF process and the
other two categories~\cite{Khoze:2002fa,Oleari:2003tc}.
Figure~\ref{fig:introduction_EW} shows
representative Feynman diagrams for these EW $\ell\ell\Pq\Pq'$ production processes.
A representative Feynman diagram for Drell--Yan production in association with two jets is
shown in Fig.~\ref{fig:introduction_DY}. This process is the dominant background in the
extraction of EW $\ell\ell\Pq\Pq'$ cross section. In what follows we designate as ``tagging jets''
the jets that originate from the fragmentation of the outgoing quarks in the EW processes shown in
Fig.~\ref{fig:introduction_EW}.

The study of these processes establishes an important foundation for the more general
study of vector boson fusion processes, of relevance for Higgs boson searches and
for measurements of electroweak gauge couplings and vector-boson scattering.
The VBF Higgs boson production in proton-proton (pp) collisions at the LHC
has been extensively investigated~\cite{atlasptdr,Bayatian:942733} as a way to
discover the particle and measure its
couplings~\cite{Zeppenfeld:2000td,Belyaev:2002ua,Duhrssen:2004cv}.
Recent searches by the ATLAS and CMS collaborations for the standard model (SM) Higgs boson
include analyses of the VBF final states~\cite{:2012gk,Chatrchyan:1471016}.

In particular, the study of the processes shown in Fig.~\ref{fig:introduction_EW}  can improve our
understanding of the selection of tagging jets
as well as that of vetoing additional parton radiation between
forward-backward jets in VBF
searches~\cite{Rainwater:1998kj,Plehn:1999xi,Rainwater:1999sd,Kauer:2000hi}.
The measurement of the electroweak production of the Zjj final state is also
a precursor to the measurement of elastic vector boson pair scattering at high energy,
an important physics goal for future analyses of LHC data.

In this work we measure the cross section for electroweak Z boson production in association
with two jets in pp collisions at a center-of-mass energy of 7\TeV, where the Z boson decays into
$\Pgmp\Pgmm$ or $\Pep\Pem$, using a data sample collected in 2011 by the Compact Muon Solenoid
(CMS) experiment with an integrated luminosity of
5.1\fbinv for the $\Pgmp\Pgmm$ mode and 5.0\fbinv for the $\Pep\Pem$ mode.
We extract the cross section under the assumption that the theory describes correctly the shape of the
kinematical distributions of the dominant background from Drell--Yan production in association with two jets

The signal-to-background ratio for the cross section measurement is small.
In order to confirm the presence of a signal, two methods of signal extraction are employed
and two different jet algorithms are used.
While providing a similar performance, these two types of jet algorithms use different
methods to combine the information from the subdetectors, different energy corrections,
and different methods to account for the energy from the additional minimum-bias events
(pileup).

In a separate study, measurements of the hadronic activity in Drell--Yan events are
presented. These include the level of hadronic activity in the rapidity interval between the
two tagging jets and the properties of multi-jets in events with a Z boson.

The plan of the paper is as follows: in Section~\ref{sec:cmsdetrecsim} we describe the
CMS detector, reconstruction, and event simulation; in Section~\ref{sec:selection} we
discuss the event selections;
Sections~\ref{sec:hadronic} and~\ref{sec:radpat} are devoted
to the study of the hadronic activity in Drell--Yan events;
in Section~\ref{sec:xsection}
we present the measurement of the cross section for the EW Zjj production; finally,
in Section~\ref{sec:conclusions} we summarize our main results.

\section{CMS detector, reconstruction, and event simulation \label{sec:cmsdetrecsim}}
A detailed description of the CMS detector can be found in Ref.~\cite{:2008zzk}.
The CMS experiment uses a right-handed coordinate system, with the origin at the nominal interaction
point, the $x$ axis pointing to the center of the LHC ring, the $y$ axis pointing up,
and the $z$ axis along the counterclockwise-beam direction as viewed from above.
The polar angle $\theta$ is measured from the positive $z$ axis and the azimuthal angle $\phi$ is
measured in the $x$-$y$ plane. The pseudorapidity $\eta$ is defined as $-\ln[\tan(\theta/2)]$,
which equals the rapidity $y = \ln[ (E+p_z) / (E-p_z)]$ for massless particles.

The central feature of the CMS apparatus is a superconducting solenoid of 6\unit{m} internal diameter
providing a field of 3.8\unit{T}. Within the field volume are a silicon pixel and strip tracker, a crystal
electromagnetic calorimeter (ECAL), and a brass/scintillator hadron calorimeter (HCAL)
providing coverage for pseudorapidities $\abs{\eta} < 3$.
The forward calorimeter modules extend the coverage of hadronic jets up to $\abs{\eta} < 5$.
Muons are measured in gas-ionization detectors embedded in the steel magnetic flux return yoke.

The first level (L1) of the CMS trigger system, composed of custom hardware processors,
uses information from the calorimeters and the muon detectors
to select the most interesting events.
The high-level-trigger
processor farm further decreases the event rate from
$\sim$100\unit{kHz} of L1 accepts to a few hundred Hz, before data storage.

Muons are reconstructed~\cite{muon} by fitting trajectories based on hits in the silicon
tracker and the muon system. Electrons are reconstructed~\cite{electron} from clusters of energy
deposits in the ECAL matched to tracks in the silicon tracker.

Two different types of jets are used in the analysis: jet-plus-track (JPT)
and particle-flow (PF) jets~\cite{jme10}.
The JPT jets are reconstructed calorimeter jets whose energy response and resolution are improved
by incorporating tracking information according to the JPT algorithm~\cite{JME-09-002}.
Calorimeter jets are first reconstructed from energy deposits
in the calorimeter towers clustered with the anti-\kt jet algorithm~\cite{Cacciari:2005hq,Cacciari:2008gp} with
a distance parameter of 0.5. Charged--particle tracks are associated with each jet,
based on the spatial separation in $\eta$-$\phi$ between the jet axis and the track momentum
vector
measured at the interaction vertex. The associated tracks are projected onto the surface of
the calorimeter and classified as in-cone tracks if they point within the jet cone around the jet axis.
The tracks bent out of the jet cone due to the magnetic field are classified as out-of-cone tracks.
The momenta of the charged tracks are used to improve the measurement of the energy of the
associated calorimeter jet. For in-cone tracks the expected average energy deposition in the
calorimeters is subtracted and the energy of the tracks (assuming that they are charged pions)
is added to the jet energy. For out-of-cone tracks the energy of the tracks is added directly
to the jet energy. The direction of the jet is re-calculated with the tracks.
As a result of the JPT algorithm, both the energy and the direction of the jet are improved.

The CMS particle flow algorithm~\cite{PFT-09-001,PFT-10-002} combines the information
from all relevant CMS sub-detectors to identify and reconstruct particle candidates in the event:
muons, electrons, photons, charged hadrons, and neutral hadrons. Charged hadrons are reconstructed
from tracks in the tracker. Photons and neutral hadrons are reconstructed from
energy clusters in the ECAL and HCAL, respectively, that are separate from the extrapolated position
of tracks. A neutral particle overlapping with charged particles in the
calorimeters is identified from a calorimeter energy excess with respect to the sum of the associated
track momenta. Particle flow jets (PF jets) are reconstructed using the anti-\kt jet algorithm
with a distance parameter of 0.5, clustering particles identified by the particle flow algorithm.

The signal process for this analysis is the electroweak production of a dilepton pair in
association with  two jets (EW $\ell\ell \pj\pj $, $\ell = \Pe,\; \mu$).
It is simulated with \MADGRAPH version 5 ~\cite{Maltoni:2002qb,Alwall:2011uj}
interfaced with \PYTHIA6.4.25~\cite{Sjostrand:2006za} for parton showering (PS) and hadronization.
The CTEQ6L1 parton distribution functions~\cite{Pumplin:2002vw} are used in the event
generation by \MADGRAPH.
The electroweak $\Pp\Pp\to \ell\ell \pj\pj $ processes in \MADGRAPH
include $\PW\cPZ$ production where the $\PW$ boson decays into two quarks and
$\cPZ\cPZ$  production where one of the $\cPZ$ bosons decays into two quarks. The requirement
$m_{\pj\pj} >120\GeV$ applied at the \MADGRAPH generation level reduces the contribution from
these processes to a negligible level in the defined signal phase space.
For the leading order generators, j stands for partons. For next-to-leading order
calculations, a jet algorithm is applied to the final state partons and j stands for the
parton jets.

Background $\cPZ$+jets (labeled DY $\ell\ell$jj) and ditop
($\ttbar$) processes are generated
with \MADGRAPH via a matrix element (ME) calculation that includes up to four jets at parton
level. The ME and parton shower
(ME-PS) matching is
performed following the ktMLM prescription~\cite{Alwall:2011uj}.
The generation of the DY $\ell\ell \pj\pj $ background does not include the electroweak
production of the $\cPZ$ boson with two jets. The diboson production processes $\PW\PW$,
$\PW\cPZ$, and $\cPZ\cPZ$ are generated with \PYTHIA.

The \MCFM program ~\cite{Campbell:2010ff} is also used for the evaluation of the
theoretical uncertainty of the DY $\ell\ell \pj\pj $ background predictions.
The dynamic scale $\mu_{0}=\sum_{i=1}^{n} \pt^{i}$ with $n$ final state particles
(partons, not jets; $n=4,5$) is used with the QCD factorization and renormalization scales
set equal, $\mu_{\mathrm{F}} = \mu_{\mathrm{R}} = \mu_{0}$.

Generated events are processed through the full CMS detector simulation based on
\GEANTfour~\cite{Agostinelli:2002hh,Allison:2006ve}, followed by a detailed trigger emulation,
and the standard event reconstruction. Minimum-bias events are superimposed upon the hard
interaction to simulate the effects of additional interactions per beam crossing (pileup).
The multiplicity distribution of the pileup events in the simulation is
matched with that observed in data.
The \PYTHIA parameters for the underlying event were set according to the
Z2 tune~\cite{Field:2011iq}.

The signal cross section per lepton flavor, at next-to-leading order (NLO),
is calculated to be $\sigma _\mathrm{NLO}$(EW $\ell\ell$jj) = 166\unit{fb}.
The calculation is carried out with the \textsc{vbfnlo} program~\cite{Arnold:2008rz}
with the factorization and renormalization scales set to $\mu_\mathrm{R}=\mu_\mathrm{F}=90\GeV$
and with CT10 parton distribution functions~\cite{Lai:2010vv}.
The calculation is performed in the following kinematical region: a dilepton invariant mass,
$m_{\ell\ell}$ above 50\GeV, jet transverse momentum $\pt^\pj  > 25\GeV$,
jet pseudorapidity $\abs{\eta^\pj} < 4$, and dijet invariant mass $m_{\pj\pj} > 120\GeV$.
The kinematic distributions for the signal
generated by \textsc{vbfnlo} at leading order agree with those produced by the \MADGRAPH generator.

The interference effects between EW and DY $\ell\ell \pj\pj $ production
processes are evaluated with the \MADGRAPH, \SHERPA~\cite{Gleisberg:2008ta},
\COMPHEP~\cite{Boos:2004kh}, and \textsc{vbfnlo} programs by the authors of these
programs and were found to be negligible.

\section{Event selection \label{sec:selection}}
For the muon channel, the candidate events were selected by a trigger that required the presence
of two muons. The requirement applied by the trigger on the muon transverse momenta
changed with increasing instantaneous luminosity.
As a consequence, the analyzed data sample is divided into three sets corresponding to the following
different thresholds: (i)~both muons have $\pt^{\mu} > 7\GeV$,
(ii)~$\pt^{\mu_{1}} > 13\GeV$ and $\pt^{\mu_{2}} > 8\GeV$, and
(iii)~$\pt^{\mu_{1}} > 17\GeV$ and $\pt^{\mu_{2}} > 8\GeV$.
Events in the electron channel were selected by a trigger that required the presence of two
electrons with $\pt^{\Pe_{1}} > 17\GeV$ and $\pt^{\Pe_{2}} > 8\GeV$.

Offline, the muon candidates used in the analysis are identified by an algorithm~\cite{muon}, which
starts from the tracks measured in the muon chambers, and then matches and combines them with the
tracks reconstructed in the inner tracker. Muons from the in--flight decays of hadrons and
punch-through particles are suppressed by applying a requirement
on the goodness-of-fit over the number of degrees of freedom, $\chi^2/\mathrm{dof}<10$,
of the global fit including the hits in the tracker and muon detectors.

In order to ensure a precise estimate of momentum and impact parameter, only tracks with more than
10 hits in the inner tracker and at least one hit in the pixel detector are used.
We require hits in at least two muon detectors, to ensure a
precise momentum estimate at the trigger level, and to suppress remaining background from
misidentified muon candidates. Cosmic muons are rejected by requiring a transverse impact parameter
distance to the beam spot position of less than 2\unit{mm}. These selection criteria provide an efficiency
of 96\% for prompt muons with $\pt > 20\GeV$. The efficiency is defined as a ratio where the
denominator is the number of generated muons with $\pt > 20\GeV$ within the geometrical
acceptance and the numerator is the number of those muons that pass the selection criteria
described above.

The electron candidates are required to pass a set of criteria which is 90\% efficient
for prompt electrons with $\pt > 20\GeV$ ~\cite{electron}. The electron identification
variables used in the selection are (i)~the spatial distance between the track
and the associated ECAL cluster, (ii)~the size and the shape of the shower in ECAL,
and (iii)~the hadronic leakage.
The track transverse impact parameter is used to discriminate electrons from conversions.
Tracks from conversions have, on average, a greater distance to the beam axis.
In order to reject electrons from conversions, candidates are allowed to have at most one missing
hit among those expected in the innermost tracker layers.

Electrons and muons from heavy-flavor decays and contained in hadronic jets
are suppressed by imposing a restriction on the presence of additional tracks around
their momentum direction. The additional tracks are summed in a cone of radius
$\Delta R = \sqrt{(\Delta\eta)^2+(\Delta\phi)^2} < 0.3$ around the lepton candidate.
Only tracks consistent with originating from the vertex corresponding to
the hardest proton-proton
scattering are used in the evaluation, so as to be insensitive to contributions from pileup
interactions in the same bunch crossing.
A relative isolation variable,
$I_{\text{trk}}  = \sum{\pt^{\text{trk}}} / \pt^{\ell}$, is evaluated for each lepton.

The dimuon channel selection ``$\cPZ_{\Pgm\Pgm}$'' is defined by
the following set of requirements: the two highest-$\pt$ muons
must have $\pt > 20\GeV$, $\abs{\eta} < 2.4$,
and must satisfy the muon quality criteria described above.
The muons are required to have
opposite charge and have a relative isolation of
$I_\text{trk} < 0.1$.
The dimuon invariant
mass is required to be within ${\pm}15\GeV$ of the Z boson mass $m_{\cPZ}=91.2\GeV$.

The following set of requirements define the ``$\cPZ_{\Pe\Pe}$'' dielectron selection:
the two higest-$\pt$
electrons must have $\pt > 20\GeV$, $\abs{\eta} < 2.4$, and satisfy the electron quality
criteria described previously.
The electrons are required to have opposite charge
and relative isolation criteria of $I_{\text{trk}} < 0.1$.

The dielectron invariant mass is required to be within
${\pm}20\GeV$ of $m_\cPZ$, a larger mass range than that for $m_{\Pgm\Pgm}$
since the dielectron Z-peak is wider because of electron bremsstrahlung effects
in the tracker material.

The two highest-$\pt$ leading jets in the event
with $\abs{\eta_{j}} < 4.7$ (labeled $\pj_1$ and $\pj_2$)
are selected as the tagging jets.
The selection criteria are optimized by maximizing
the signal significance defined as $N_{S}/\sqrt{N_{B}}$, where $N_{S}$ and $N_{B}$
are the number
of signal and background events passing the selection criteria, expected from the
Monte~Carlo (MC) signal
and DY samples, with an integrated luminosity of 5\fbinv.
The requirements on the momentum and pseudorapidity of the tagging jets
($\pt^{\pj_1}$, $\pt^{\pj_2}$, $\eta_\pj $),
the dijet invariant mass ($m_{\pj_{1}\pj_{2}}$), and the $\cPZ$ boson rapidity in
the rest frame of the tagging jets $y^{\ast} = y_{Z} - 0.5(y_{\pj_1}+y_{\pj_2})$
are varied in order to reach maximum signal significance. The optimized selection
criteria shown in Table~\ref{tab:names_tab}, with the corresponding selection
labels, result in an expected signal significance of about three for an integrated
luminosity of 5\fbinv, for each of the dilepton channels.

\begin{table}[ht!]
\centering
\topcaption{The optimized selection criteria with the corresponding selection labels.}
\begin{tabular}{|l|c|}
\hline
\multicolumn{2}{|c|}{Tagging jet selections} \\\hline
TJ1 & $\pt^{\pj_{1}} > 65\GeV$, $\pt^{\pj_{2}} > 40\GeV$, $\abs{\eta_\pj}<3.6$ \\
TJ2 & $m_{\pj_{1}\pj_{2}} >600\GeV$ \\\hline
\multicolumn{2}{|c|}{Z boson rapidity selection} \\\hline
YZ  & $\abs{{y^{\ast}}} < 1.2$ \\
\hline
\end{tabular}
\label{tab:names_tab}
\end{table}

The signal efficiencies for the $\cPZ_{\Pgm\Pgm}$ selection without additional
requirements, with the tagging jet requirement TJ1, with the TJ1 and the Z boson rapidity
requirement YZ, and with the TJ1, YZ, and TJ2 requirements are 0.36, 0.23, 0.17, and 0.06, respectively.
These efficiencies are valid both in the case of JPT and PF jet reconstruction.
The signal efficiencies for the dielectron channel are respectively 0.33, 0.21, 0.16,
and 0.06. The efficiency is defined as a ratio where the denominator is the number of
signal events generated by \MADGRAPH with $m_{\pj\pj}>120\GeV$ and the numerator is the
number of events that passed the selections described above.

\begin{figure}[htp]
\begin{center}
\begin{tabular}{c}
\includegraphics[width=0.45\textwidth]{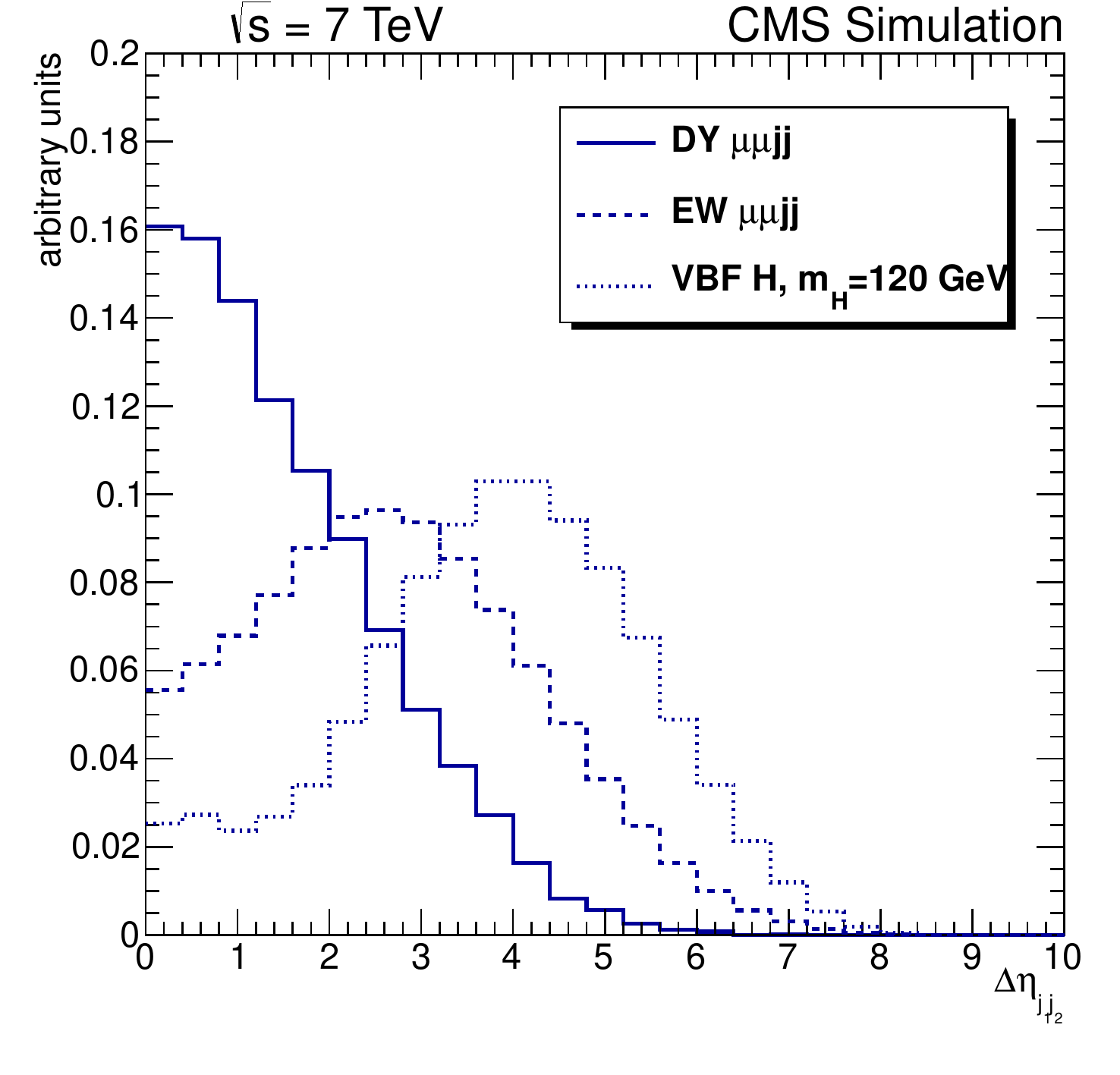}
\includegraphics[width=0.45\textwidth]{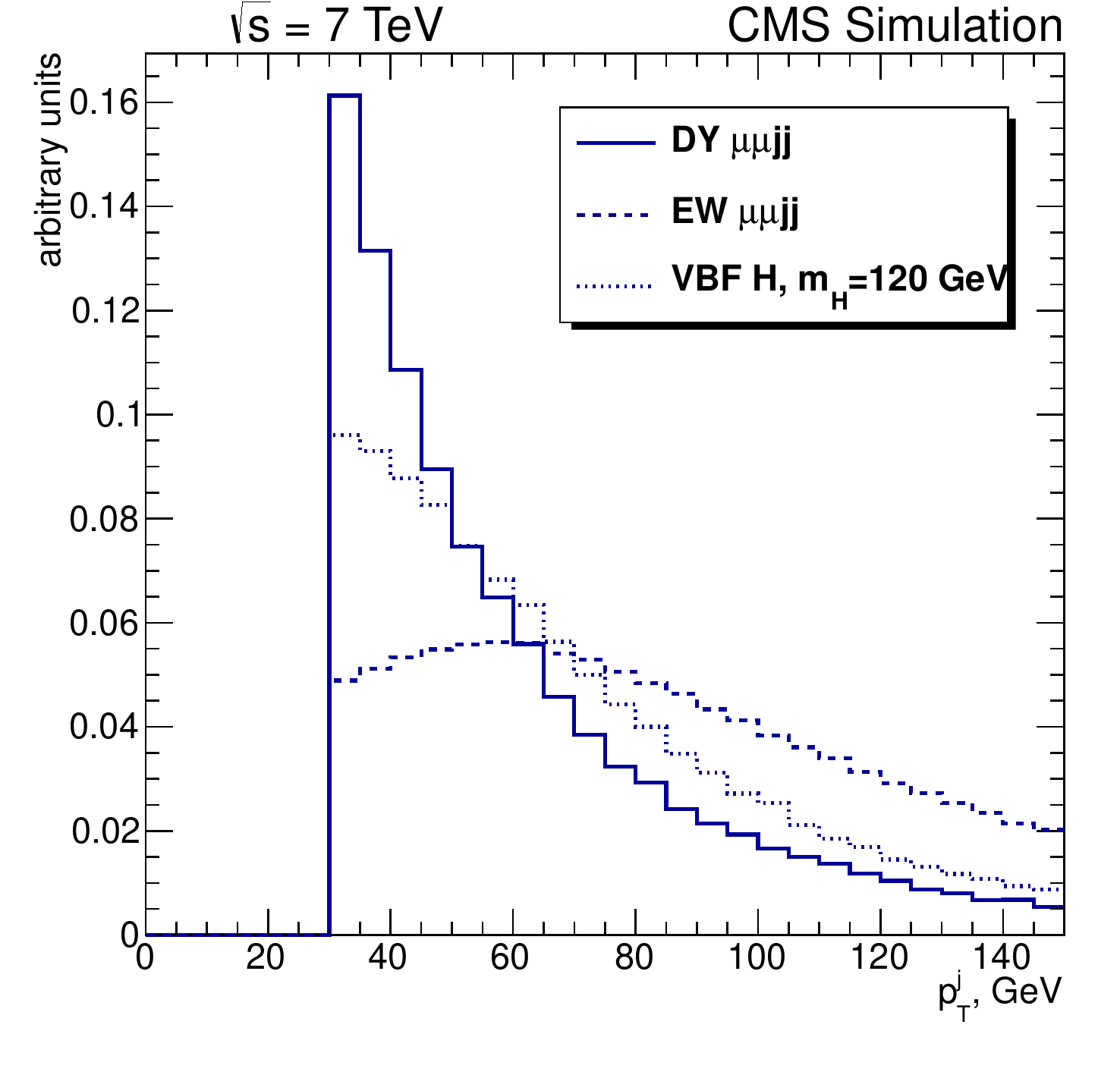}
\end{tabular}
\caption{Distribution of the absolute difference in the pseudorapidity of the tagging jets,
         $\Delta\eta_{{\pj_{1}\pj_{2}}} = \abs{\eta_{{\pj_{1}}} - \eta_{{\pj_{2}}}}$
        (left) and the tagging jet $\pt$ for both jets, $\pj_{1}$ and $\pj_{2}$ (right)
         for the DY $\Pgm\Pgm \pj\pj $, EW $\Pgm\Pgm \pj\pj $, and VBF Higgs boson production processes.}
\label{fig:kine}
\end{center}
\end{figure}

The above event selection criteria are different from those suggested for
Higgs boson searches in the VBF
channel~\cite{Rainwater:1998kj,Plehn:1999xi,Rainwater:1999sd,Kauer:2000hi}.
In particular, higher $\pt$ thresholds are used on the tagging jets,
the rapidity separation between the tagging jets is not used, and a central jet veto is not applied.
This is because the kinematics for EW Zjj production and
VBF Higgs boson production are different. The former includes
two
additional contributions, bremsstrahlung and multiperipheral processes, as shown
in Fig.~\ref{fig:introduction_EW}. These additional processes and the
interference between them lead to higher average jet transverse momenta
in comparison to
the VBF production process alone. This is due to the fact that the EW $\ell\ell$jj process involves
transversely polarized W bosons, while the main contribution to
VBF boson production involves longitudinally polarised W bosons.
Figure~\ref{fig:kine} shows the simulated
distributions of the absolute pseudorapidity difference of the two tagging jets,
$\Delta\eta_{{\pj_{1}\pj_{2}}} = \abs{\eta_{\pj_{1}} - \eta_{\pj_{2}}}$ (left), and
the tagging jets $\pt$ (right) for the DY $\Pgm\Pgm \pj\pj $, the EW $\Pgm\Pgm \pj\pj $, and
the VBF Higgs boson production processes. The $m_{\pj_{1}\pj_{2}}$ distributions for
EW $\Pgm\Pgm \pj\pj $ production and VBF Higgs boson production processes are very similar and
are not shown here.

The event selection is performed with JPT and PF jets in the dimuon channel
and with PF jets in dielectron channel.
Table~\ref{tab:selections_tab} shows the event yield after each selection step in the $\Pgmp\Pgmm$
channel. The observed and expected number of events from signal and background processes
are shown for the different selection requirements.
The two jet algorithms result in similar yields.
Table~\ref{tab:elselections_tab} shows the event yields after each selection step
with PF jets in the $\Pep\Pem$ channel.

\begin{table}[ht!]
\centering
\topcaption{Event yields in the $\Pgmp\Pgmm$ channel after each selection step for the data,
the signal Monte Carlo and the backgrounds.
         The expected contributions from the signal and background processes are evaluated
         from simulation, for 5\fbinv of integrated luminosity.}
\begin{tabular}{|c|c|c|c|c|c|c|c|c|}
\hline
Selection              & Jet type  & Data   &  EW $\ell\ell \pj\pj$  & DY $\ell\ell \pj\pj$
&  $\ttbar$  &    WW     &     WZ    &     ZZ    \\
\hline
\hline
$\cPZ_{\Pgm\Pgm}$  &           & $1.7\times10^{6}$
                                   & 460
                                   & $1.7\times10^{6}$
                                   & 1400
                                   &  300
                                   & 1300
                                   &  850    \\\hline
requirement TJ1    &    JPT   &     25000         &    290          &    26000       &    690    &     5.2   &    180     & 120      \\
                   &    PF   &      26000         &    280          &    26000       &    680    &     5.3   &    170     & 110     \\\hline
requirement YZ    &    JPT   &    15000          &    210          &    16000       &    590    &     3.4   &    98      & 83    \\
                   &    PF   &     16000          &    200          &    16000       &    580    &     3.4   &    93      & 76     \\\hline
requirement TJ2    &    JPT   &    600            &    74           &    600         &    14     &     0     &    2.2     & 1.3    \\
                   &    PF   &     640            &    72           &    610         &    14     &     0     &    2.4     & 1.2     \\\hline
\end{tabular}

\label{tab:selections_tab}
\end{table}

\begin{table}[ht!]
\centering
\topcaption{Event yields in the $\Pep\Pem$ channel after each selection step for the data,
the signal Monte Carlo and the backgrounds.
         The expected contributions from the signal and background processes are evaluated
         from simulation, for 5\fbinv of integrated luminosity.}
\begin{tabular}{|c|c|c|c|c|c|c|c|}
\hline
selection              & data   &  EW $\ell\ell \pj\pj $  & DY $\ell\ell \pj\pj $
&  $\ttbar$  &    WW     &     WZ    &     ZZ    \\
\hline
\hline
$\cPZ_{\Pe\Pe}$  & $1.5\times 10^{6}$ & 410 & $1.5\times 10^{6}$ & 1600& 340 & 1100 & 720 \\
requirement TJ1        & 24000 & 270 & 23000 & 880 & 6.0 & 150 & 97 \\
requirement YZ         & 15000 & 200 & 15000 & 760 & 3.7 &  90 & 68 \\
requirement TJ2        &   560 &  67 &   550 &  17 & 0.3 &  2.5 & 1.0 \\\hline
\end{tabular}
\label{tab:elselections_tab}
\end{table}

The uncertainty on the estimation of the dominant DY $\ell\ell \pj\pj $ background from simulation
is comparable with the expected number of signal events. The signal can therefore only be extracted by
analyzing the distributions that are most sensitive to the difference between the signal and
backgrounds.

Distributions for data and simulation after the $\cPZ_{\Pgm\Pgm}$ selection
and jet tagging requirement TJ1 are shown in Figs. \ref{fig:mumu_ptj1_ptj2} and \ref{fig:mumu_deta_ptz}. In these and the following figures the histograms with the labels ``DY'' and ``ttbar'' show
the contributions from the DY $\ell\ell \pj\pj $ and $\ttbar$ processes. The labels ``WZ'', ``ZZ'',
and ``WW'' apply to the diboson production processes WZ, ZZ, and WW. The label ``EW'' shows the
contribution from the signal process, EW Zjj.

The $\pt^{\pj_{1}}$ and $\pt^{\pj_{2}}$ distributions obtained with JPT jets are shown in
Fig.~\ref{fig:mumu_ptj1_ptj2}.
The absolute difference in the pseudorapidity of the two tagging JPT jets, and the dimuon $\pt$ are shown
in Fig.~\ref{fig:mumu_deta_ptz}.
The expected contributions from the signal and background processes are evaluated from simulation.
The bottom panel in the figures shows the ratio of the data to the expected contribution of the
signal plus background together with the statistical uncertainties. The region between the two lines
with the labels JES Up and JES Down shows the $\pm 1\sigma$ uncertainty of the simulation prediction
due to the jet energy scale (JES) uncertainty.

The ratio of the data to the expected contribution of
the signal plus background is systematically below unity and outside the 1 $\sigma$ JES
uncertainty in some regions. However, it is consistent with unity within the systematic
uncertainty in the
\MADGRAPH simulation of the dominant DY $\ell\ell \pj\pj $ background. The systematic uncertainty
due to the QCD scale is expected to be between the uncertainty given by the NLO and LO calculations,
which are  8\% and 25\%, respectively, as calculated by the \MCFM program. The choice of the QCD
scale is discussed in Section~\ref{sec:systematicsjpt}.

Figures \ref{fig:mumu_ptj1_ptj2} and \ref{fig:mumu_deta_ptz}
illustrate the overall level of agreement between
data and simulation. It is evident from the figures that the signal
fraction is small; this is why the extraction of the signal requires the
special methods described in Section~\ref{sec:xsection}.

In Sections~\ref{sec:hadronic} and~\ref{sec:radpat} we describe the measurements
of the hadronic activity in the rapidity interval between the tagging jets and the measurements
of the radiation patterns in multijet events in association with a Z boson. The selected
data sample is dominated by DY $\ell\ell \pj\pj $ events which are
referred to as ``DY Zjj events'' in the following two Sections.
\begin{figure}[htp]
\begin{center}
\begin{tabular}{c}
\includegraphics[width=0.50\textwidth]{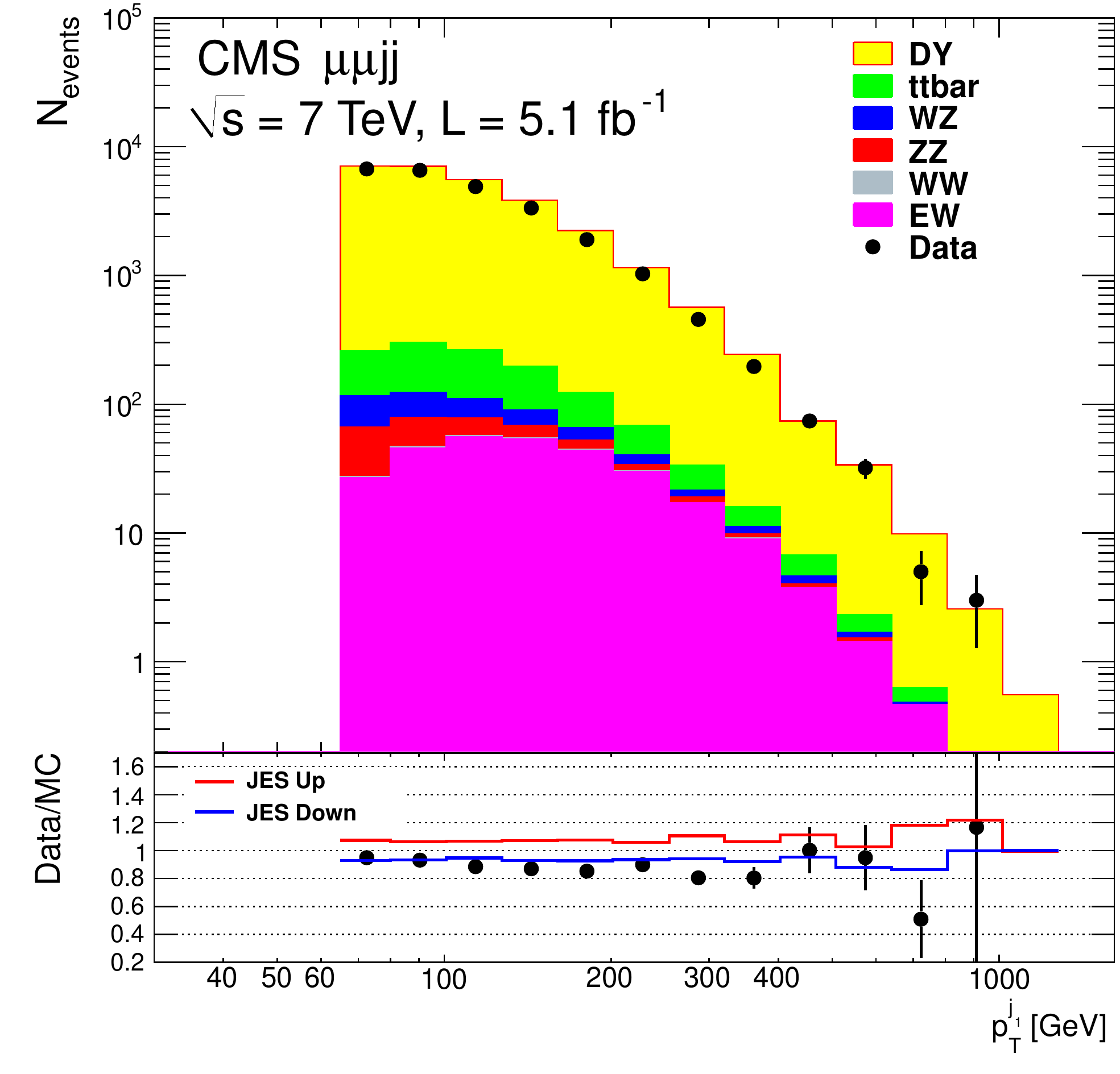}
\includegraphics[width=0.50\textwidth]{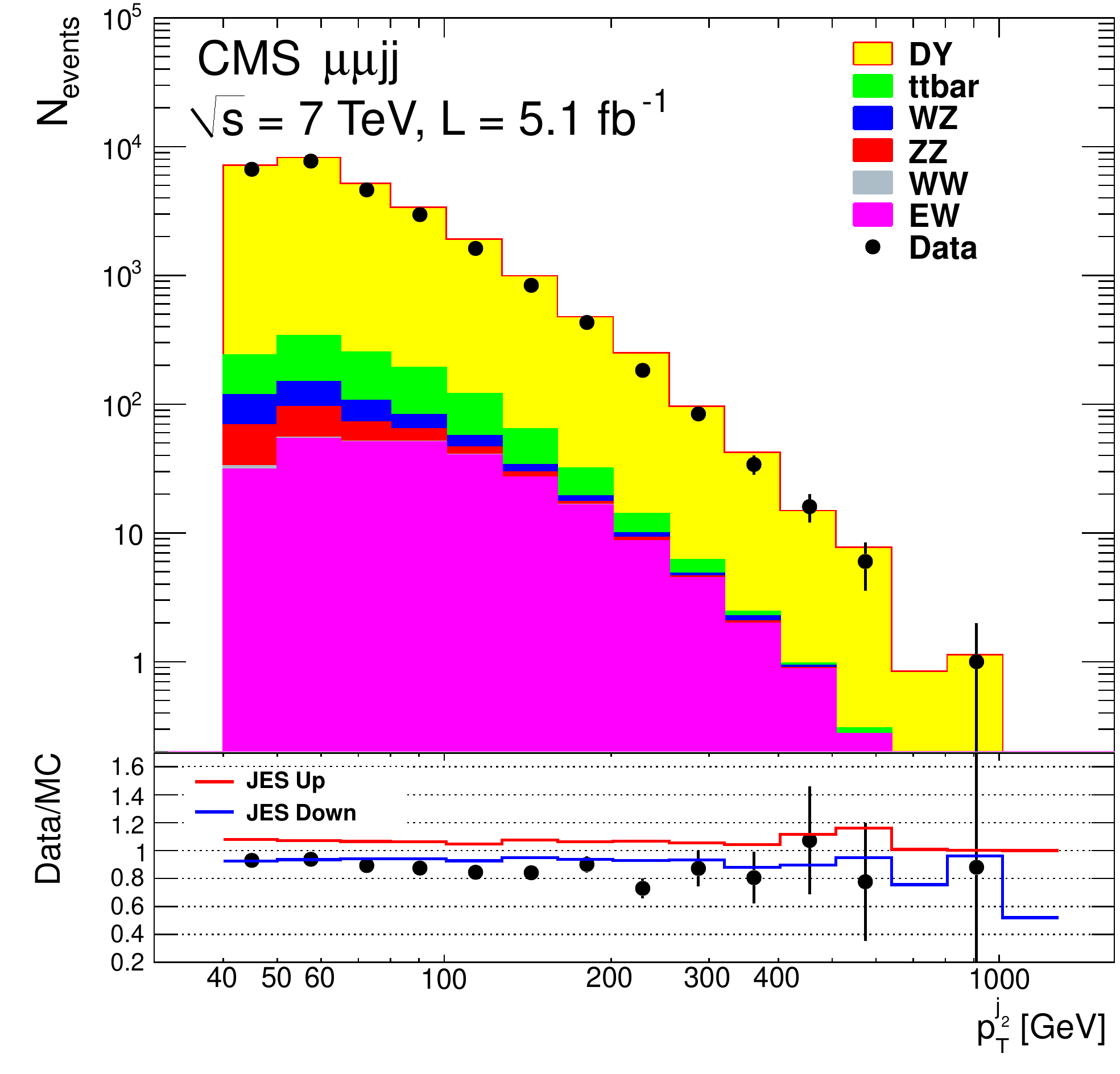}
\end{tabular}
\caption{The $\pt^{\pj_{1}}$ (left) and $\pt^{\pj_{2}}$ (right) distributions after
         applying the $\cPZ_{\Pgm\Pgm}$ selection and the jet tagging requirement TJ1.
         The expected contributions from the signal and background processes are
         evaluated from simulation. The bottom panels show the ratio of data over the expected
         contribution of the signal plus background.
	 The region between the two lines with the labels JES Up and JES Down shows the 1 $\sigma$
         uncertainty of the simulation prediction due to the jet energy scale uncertainty.
         The data points are shown with the statistical uncertainties.}
\label{fig:mumu_ptj1_ptj2}
\end{center}
\end{figure}

\begin{figure}[htp]
\begin{center}
\begin{tabular}{c}
\includegraphics[width=0.50\textwidth]{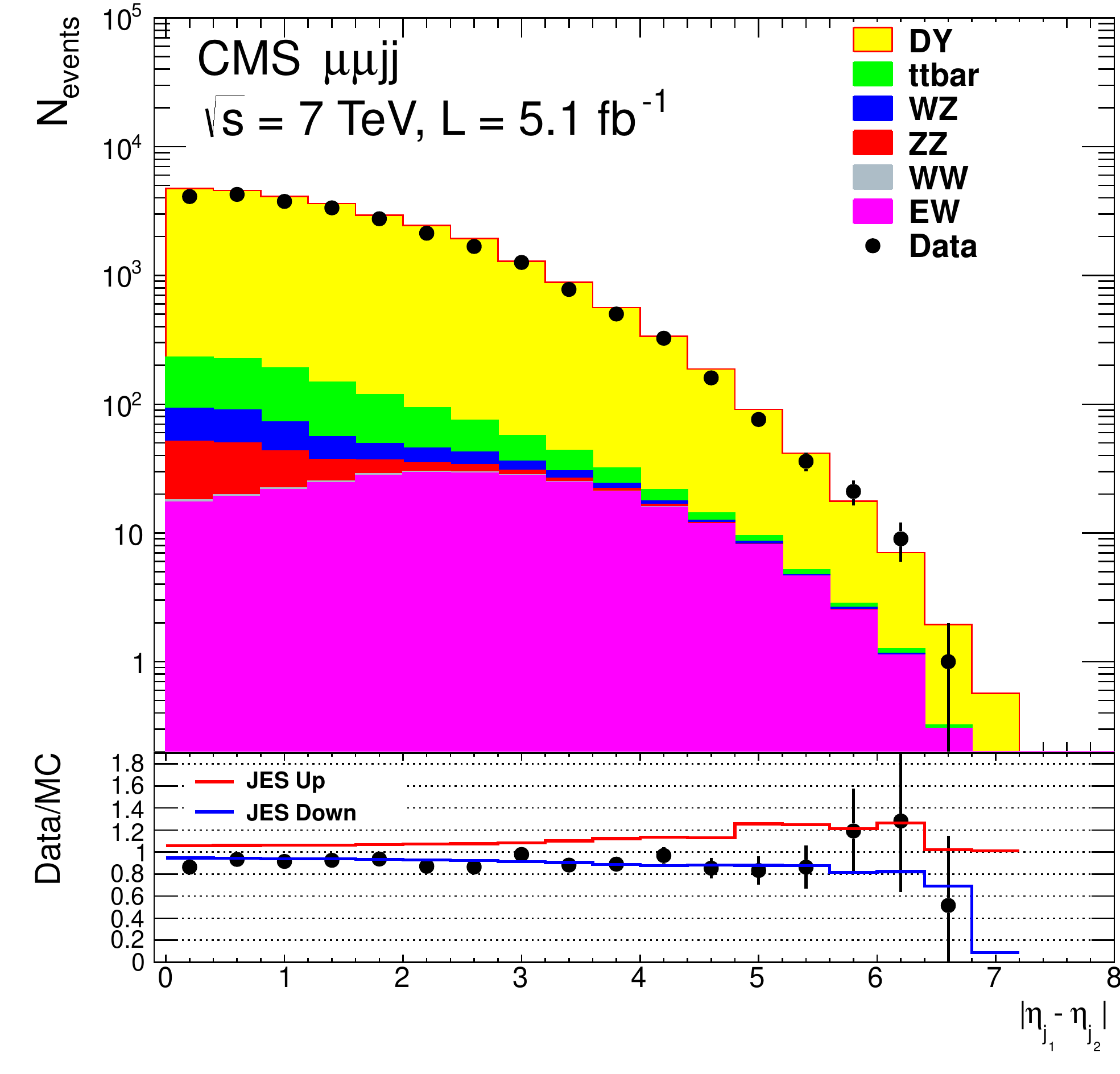}
\includegraphics[width=0.50\textwidth]{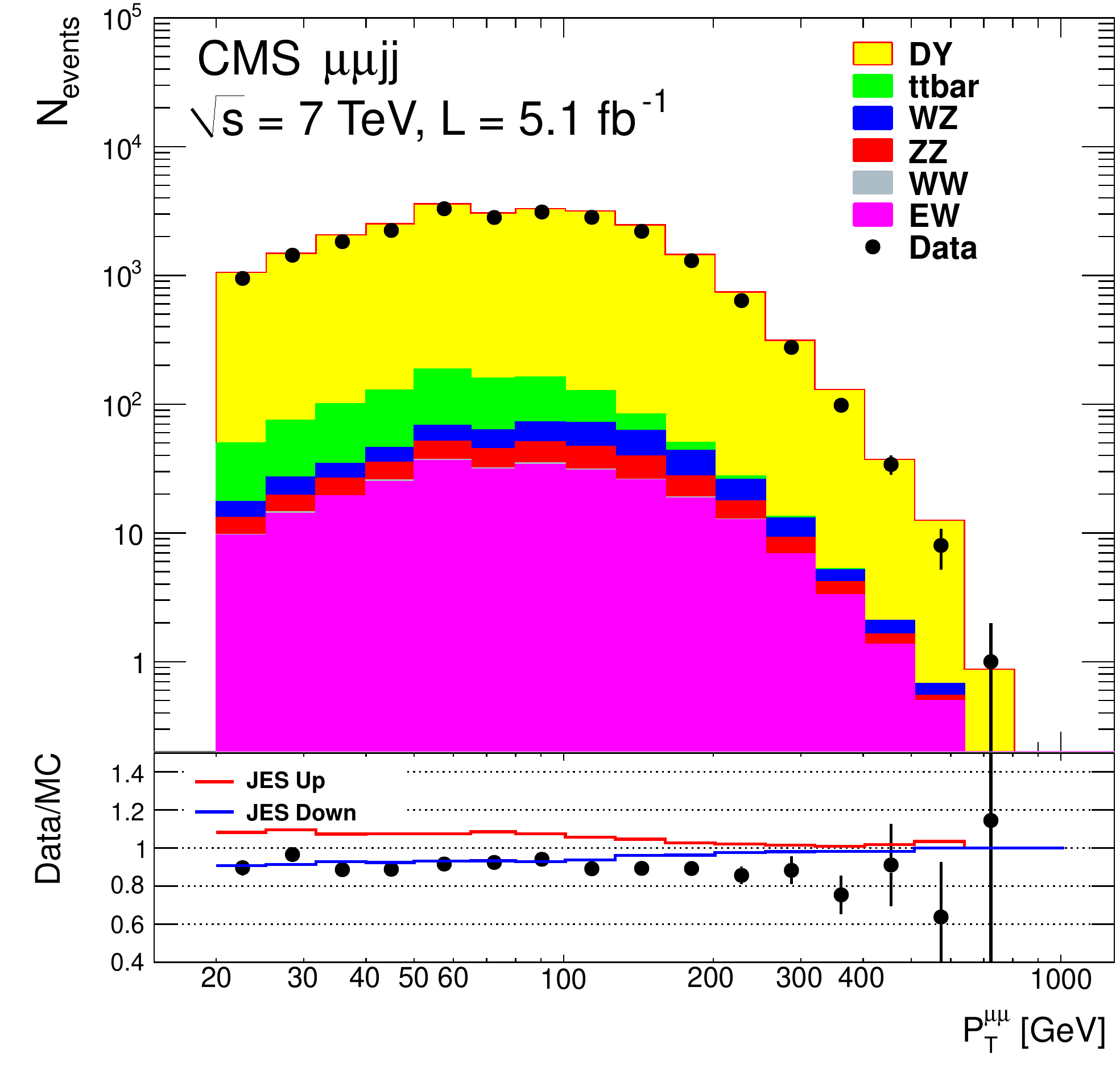}
\end{tabular}
\caption{The absolute difference in the pseudorapidity of the two tagging jets (left),
         and the dimuon $\pt$ (right) after the $\cPZ_{\Pgm\Pgm}$ selection and
         the tagging jet requirement TJ1.
         The expected contributions from the signal and background
         processes are evaluated from simulation. The bottom panels show the ratio of data
         over the expected contribution of the signal plus background.
         The region between the two lines with the labels JES Up and JES Down shows the
         1$\,\sigma$ uncertainty of the simulation prediction due to the jet energy scale
         uncertainty. The data points are shown with the statistical uncertainties.}
\label{fig:mumu_deta_ptz}
\end{center}
\end{figure}

\section{Hadronic activity in the rapidity interval between tagging jets\label{sec:hadronic}}
A veto on the hadronic activity in the rapidity interval between the VBF tagging jets
has been proposed~\cite{Rainwater:1998kj,Plehn:1999xi,Rainwater:1999sd,Kauer:2000hi}
as a tool to suppress backgrounds in the searches for a Higgs boson produced in VBF.
In the following, a study of the hadronic activity in this
rapidity interval is presented.
Although a veto is not used on the hadronic activity
to select the EW $\ell\ell \pj\pj $
process, the studies provided in this Section and in Section~\ref{sec:radpat} can be considered
as a test of the agreement between the data and the simulation for the dominant DY $\ell\ell \pj\pj $
background. The data sample is selected with the $\cPZ_{\Pgm\Pgm}$ and $\cPZ_{\Pe\Pe}$
requirements described in Section~\ref{sec:selection}. The requirements on the jets are described
in this Section and in Section~\ref{sec:radpat}.

\subsection{Central hadronic activity measurement using jets \label{sec:cjv}}

The hadronic activity in the rapidity interval between the tagging jets
is studied
as a function of the pseudorapidity separation between the tagging jets,
the $\pt$ threshold of the tagging jets, and the dijet invariant mass, $m_{\pj_{1}\pj_{2}}$. The
hadronic activity is measured through the efficiency of the central jet veto,
defined as the fraction of selected events
with no third jet ($\pj_{3}$) with $\pt^{\pj_{3}} > 20$\GeV in the
pseudorapidity interval between the tagging jets:

\begin{gather}
\eta_{\text{min}}^{\text{tag jet}}+0.5 < \eta _{\pj_{3}} < \eta_{\text{max}}^{\text{tag jet}}-0.5
       \qquad\text{and}\qquad\abs{\eta_{\pj_{3}}} < 2.0,
\end{gather}

where $\eta_\text{min}^\text{tag jet}$ ($\eta_\text{max}^\text{tag jet}$) is the minimal (maximal)
pseudorapidity of the tagging jet. The central jets from pileup interactions are
suppressed with the tracker information.

Tables~\ref{tab:cjv1_tab} and~\ref{tab:cjv2_tab} show the efficiencies measured
from the data and those obtained from the \MADGRAPH DY $\ell\ell \pj\pj $ simulation for the
different requirements on the $\pt$ of the tagging jets, the pseudorapidity separation between
them, and their invariant mass. The measured efficiency is shown
with the statistical uncertainties. The contribution of the EW Zjj, $\ttbar$, and diboson
processes is not subtracted from the data measurements since it does not change the
measured efficiency within the uncertainties.
The efficiencies shown in Table~\ref{tab:cjv2_tab} are evaluated for the pseudorapidity
interval

\begin{gather}
\eta_{\text{min}}^{\text{tag jet}} < \eta _{\pj_{3}} < \eta_{\text{max}}^{\text{tag jet}}
       \qquad\text{and}\qquad\abs{\eta_{{\pj_{3}}}} < 2.0 .
\end{gather}

\begin{table}[ht!]
\centering
\topcaption{Efficiency of the central jet veto with $\pt^{\pj_{3}} > 20$\GeV for
         three different selections on the tagging jets for a pseudorapidity separation of
         $\Delta \eta _{\pj_{1}\pj_{2}}> 3.5$ measured in data and predicted by the
\MADGRAPH simulation. The quoted uncertainty is statistical only.
}
\begin{tabular}{|c|c|c|c|}
\hline
 $\pt^{\pj_{1}(\pj_{2})}$  & $ >$25\GeV
                   & $>$35\GeV
                   & $>$45\GeV  \\
\hline
data               & $0.78\pm0.01$  & $0.68\pm0.01$   & $0.63\pm0.02$  \\
simulation         &      0.80      &     0.71        &    0.66        \\
\hline
\end{tabular}

\label{tab:cjv1_tab}
\end{table}

\begin{table}[ht!]
\centering
\topcaption{Efficiency of the central jet veto with $\pt^{\pj_{3}} > 20$\GeV and
         $\pt^{\pj_{1}(\pj_{2})} > 30$\GeV for three different selections for
         $\Delta \eta _{\pj_{1}\pj_{2}}$ with and without the selection on
         $m_{\pj_{1}\pj_{2}}$, measured in data and predicted by the
\MADGRAPH simulation. The quoted uncertainty on the data efficiency is only statistical.}
\begin{tabular}{|c|c|c|c|}
\hline
$\Delta \eta _{\pj_{1}\pj_{2}}$   & $>$2.5
                   & $>$3.5
                   & $>$4.5 \\
\hline
data               & $0.71\pm0.01$  & $0.68\pm0.01$   & $0.66\pm0.02$  \\
simulation         &      0.73      &     0.71        &    0.67        \\\hline
\hline
\multicolumn{4}{|c|}{with $m_{\pj_{1}\pj_{2}} > 700$\GeV selection}           \\
\hline
data               & $0.56\pm0.03$  & $0.58\pm0.03$   & $0.62\pm0.04$  \\
simulation         &      0.56      &     0.57        &    0.58        \\
\hline
\end{tabular}

\label{tab:cjv2_tab}
\end{table}

The veto efficiencies obtained from data and
the \MADGRAPH simulation are in good agreement.

\subsection{Central hadronic activity measurement with track jets \label{sec:trackjets}}

As the hadronic activity in the rapidity interval between the tagging jets is expected
to be small (soft) in the case of a purely
electroweak Zjj production, the contribution from any additional pileup interaction in the event
needs to be avoided or carefully subtracted.
For this reason, an additional study of the interjet hadronic activity
is performed using only charged tracks that clearly originate from the hard-scattering
vertex in the event.

For this study a collection of tracks is built with reconstructed high-purity
tracks~\cite{CMS-PAS-TRK-10-005} with $\pt > 300\MeV$
that are uniquely associated with the main primary vertex in the event.
Tracks associated with the two leptons or with the tagging jets are not included.
The association between the tracks and the reconstructed primary
vertices is carried out by minimizing the longitudinal distance $d_z$(PV)
between the primary vertex (PV) and the point of closest approach of the track helix to that PV.
The association is required to satisfy $d_z\mathrm{(PV)}<2\unit{mm}$ and
 $d_z\mathrm{(PV)}<3\delta d_z$(PV), where $\delta d_z$(PV) is the uncertainty on $d_z$(PV).
The main primary vertex in the event is chosen to be that with the largest scalar sum
of transverse momenta, for all tracks used to reconstruct it.

A collection of ``soft track jets'' is built
by clustering the tracks with the anti-\kt clustering algorithm~\cite{Cacciari:2008gp}
with a distance parameter of 0.5. The use of track jets represents a clean and well
understood method~\cite{CMS-PAS-JME-10-006} to reconstruct jets with energy
as low as a few \GeVns{}.
Crucially, these jets are not affected by pileup because of the association of
their tracks
with the hard-scattering vertex~\cite{CMS-PAS-JME-08-001}.

\begin{figure}[hbtp]
  \begin{center}
    \includegraphics[width=7.5cm]{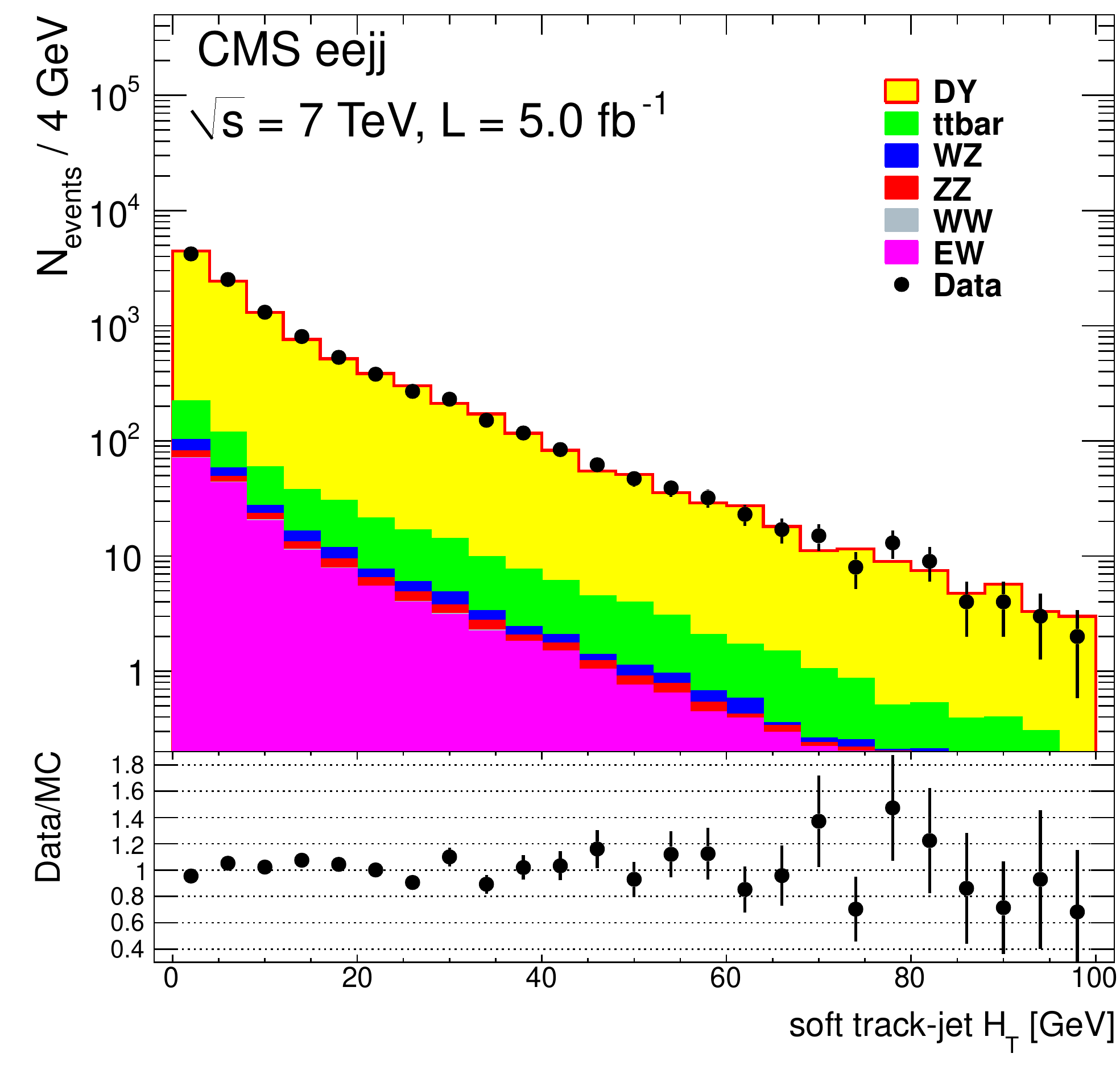}
    \includegraphics[width=7.5cm]{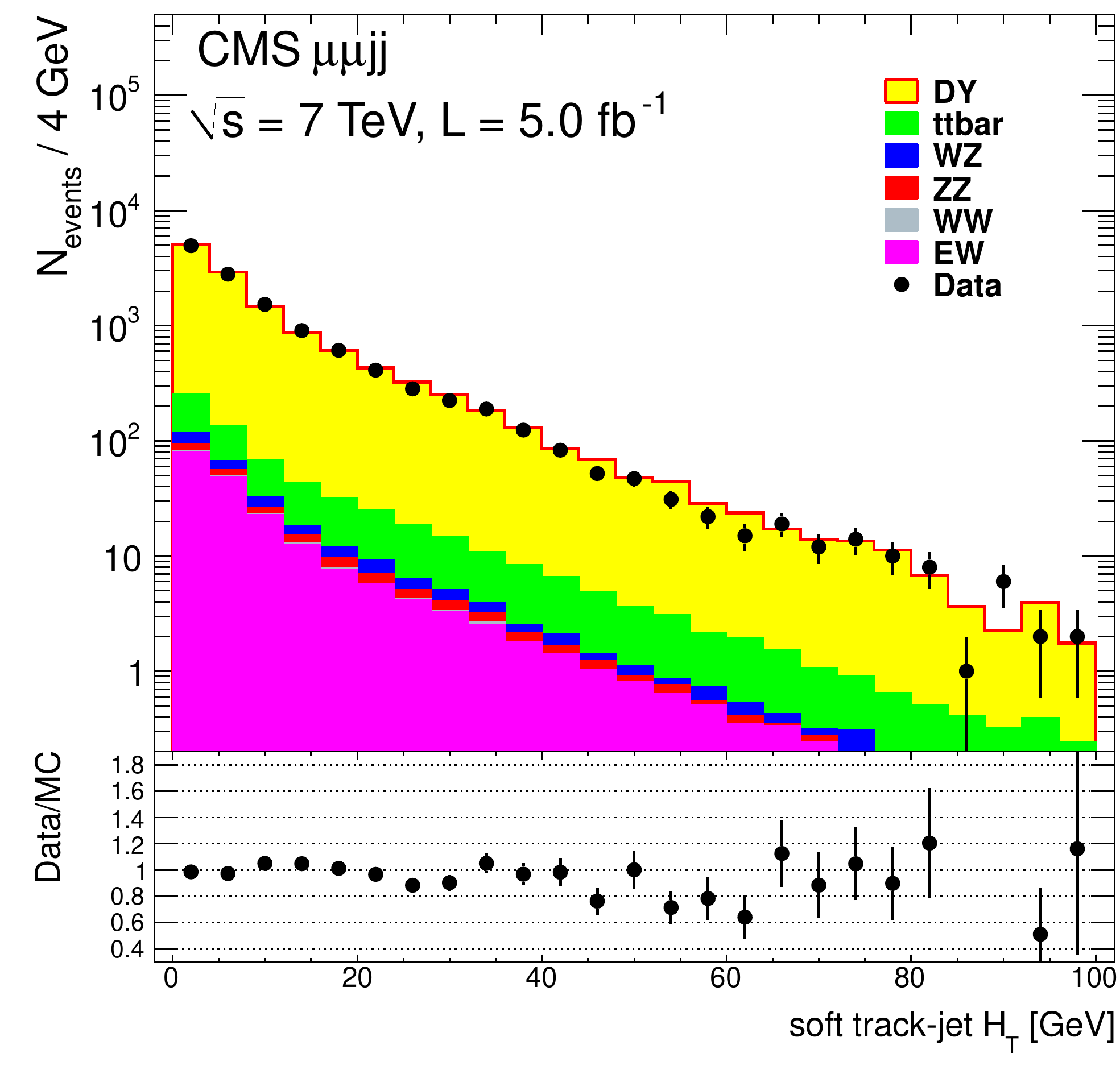}
    \caption{The $\HT$ distribution of the three leading soft track jets in the
             pseudorapidity interval between the tagging jets with
             $\pt^{\pj_{1},\pj_{2}} > 65,\,40\GeV$ in DY Zjj events for dielectron (left)
             and dimuon (right) channels. The bottom panels show the corresponding data/MC
             ratios. The data points are shown with the statistical uncertainties.}
    \label{fig:H_T}
  \end{center}
\end{figure}

For the purpose of studying the central hadronic activity between the tagging jets,
only soft track jets with pseudorapidity
$\eta^\text{tag jet}_\text{min}+0.5 < \eta < \eta^\text{tag jet}_\text{max}-0.5 $ are
considered.
The scalar sum ($\HT$) of the transverse momenta of up to three soft track jets is
used as a monitor of the hadronic activity in the rapidity interval between the two
jets. The soft $\HT$ distribution is shown in Fig.~\ref{fig:H_T} for
DY Zjj events  for $\pt^{\pj_{1},\pj_{2}} > 65,\,40\GeV$.
The expectations from the simulation for the hadronic activity between the tagging jets are
in good agreement with the data.

\begin{figure}[hbtp]
  \begin{center}
    \includegraphics[width=7.5cm]{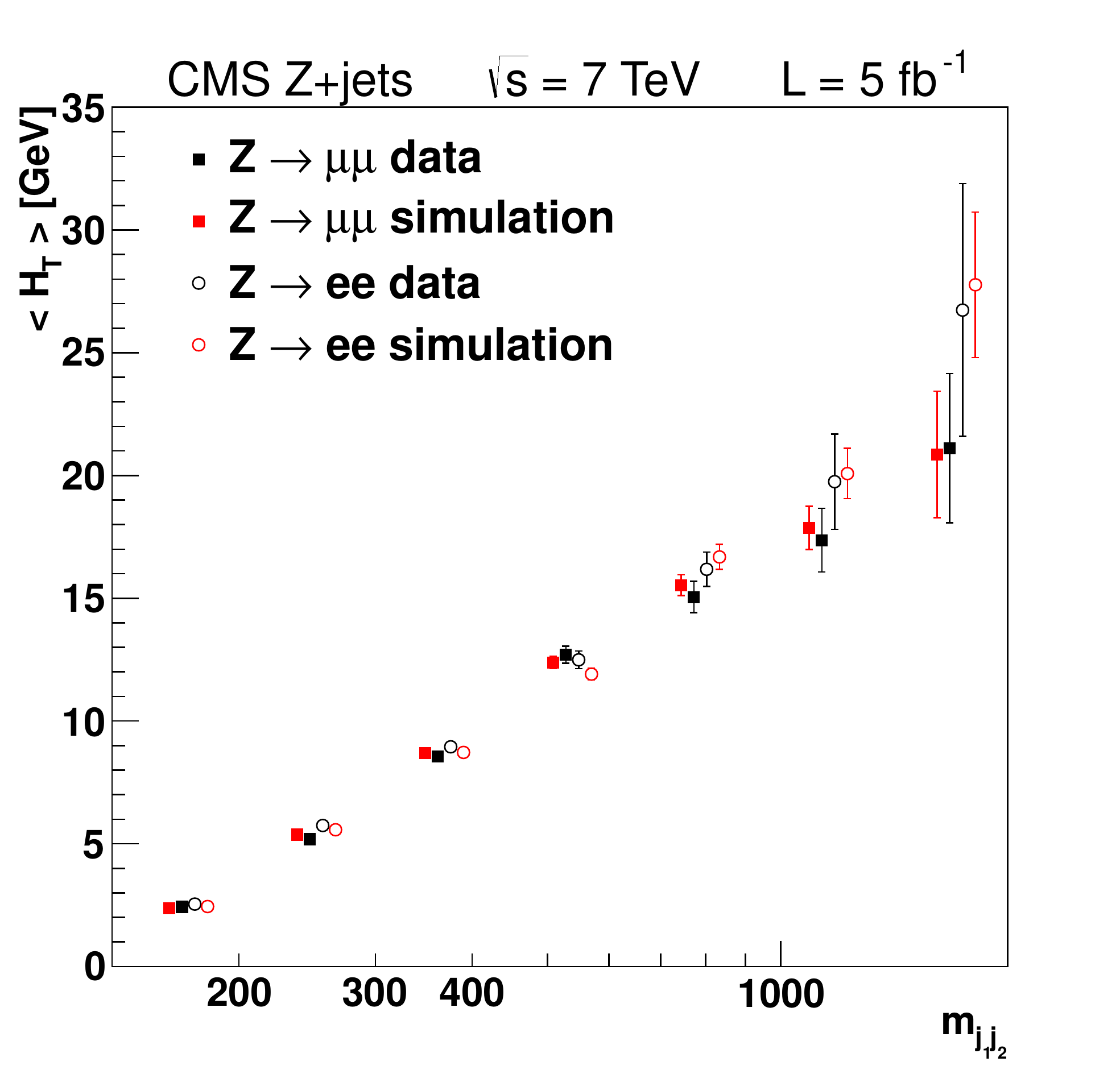}
    \includegraphics[width=7.5cm]{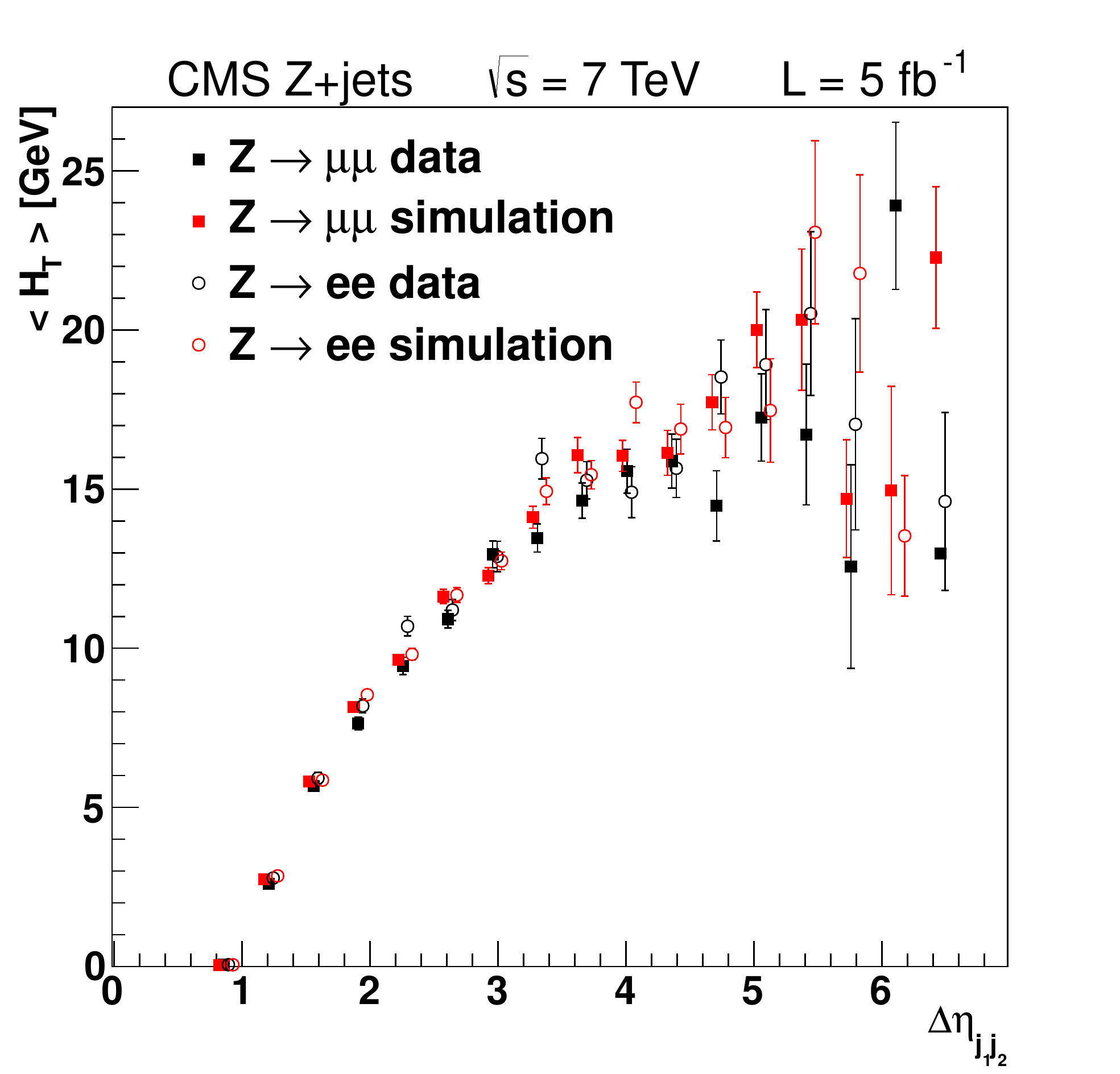}
    \caption{Average $\HT$ of the three leading soft track jets in the
             pseudorapidity gap between the tagging jets for
             $\pt^{\pj_{1},\pj_{2}} > 65,\,40\GeV$ as a function of the dijet
             invariant mass (left) and the dijet $\Delta\eta_{\pj_{1}\pj_{2}}$ separation
             (right) for both the dielectron and dimuon channels
              in DY Zjj events.
              The data points and the points from simulation are shown with
              the statistical uncertainties.}
    \label{fig:HT_vs}
  \end{center}
\end{figure}

The evolution of the average $\HT$ for DY Zjj jets events as a function of the dijet
invariant mass $m_{\pj_{1}\pj_{2}}$ and the pseudorapidity difference
$\Delta\eta_{\pj_{1}\pj_{2}}$ between the tagging jets is shown in Fig.~\ref{fig:HT_vs}.
For better visibility the symbols at each measured point are slightly displaced along
the x axis. Good agreement is observed between the simulation and the data for the
different mass and pseudorapidity intervals.

\section{Measurements of the radiation patterns in multijet events
         in association with a Z boson \label{sec:radpat}}

In hard multijet events in association with a Z boson, the observables referred to as
``radiation patterns" are:
\begin{itemize}
\item{the number of jets $N_\pj$;}
\item{the total scalar sum ($\HT$) of jets with $\abs{\eta}<4.7$;}
\item{the difference in the pseudorapidity, $\Delta\eta_{\pj_{1}\pj_{2}}$, between the two most
      forward-backward jets (which are not necessarily the two highest--$\pt$ jets);}
\item{the cosine of the azimuthal angle difference,
      $\cos \abs{\phi_{\pj_1} - \phi_{\pj_2}} = \cos \Delta\phi_{\pj_1 \pj_2}$,
      between the two most forward-backward jets.}
\end{itemize}

These observables are investigated following the prescriptions and suggestions in
Ref.~\cite{Binoth:2010ra}, where the model dependence is estimated by comparing the
predictions from \MCFM~\cite{Campbell:2010ff}, \PYTHIA,
\ALPGEN~\cite{Mangano:2002ea}+\PYTHIA, and the \textsc{hej}~\cite{Andersen:2011hs} programs.

\begin{figure}[hbtp]
  \begin{center}
    \includegraphics[width=7.5cm]{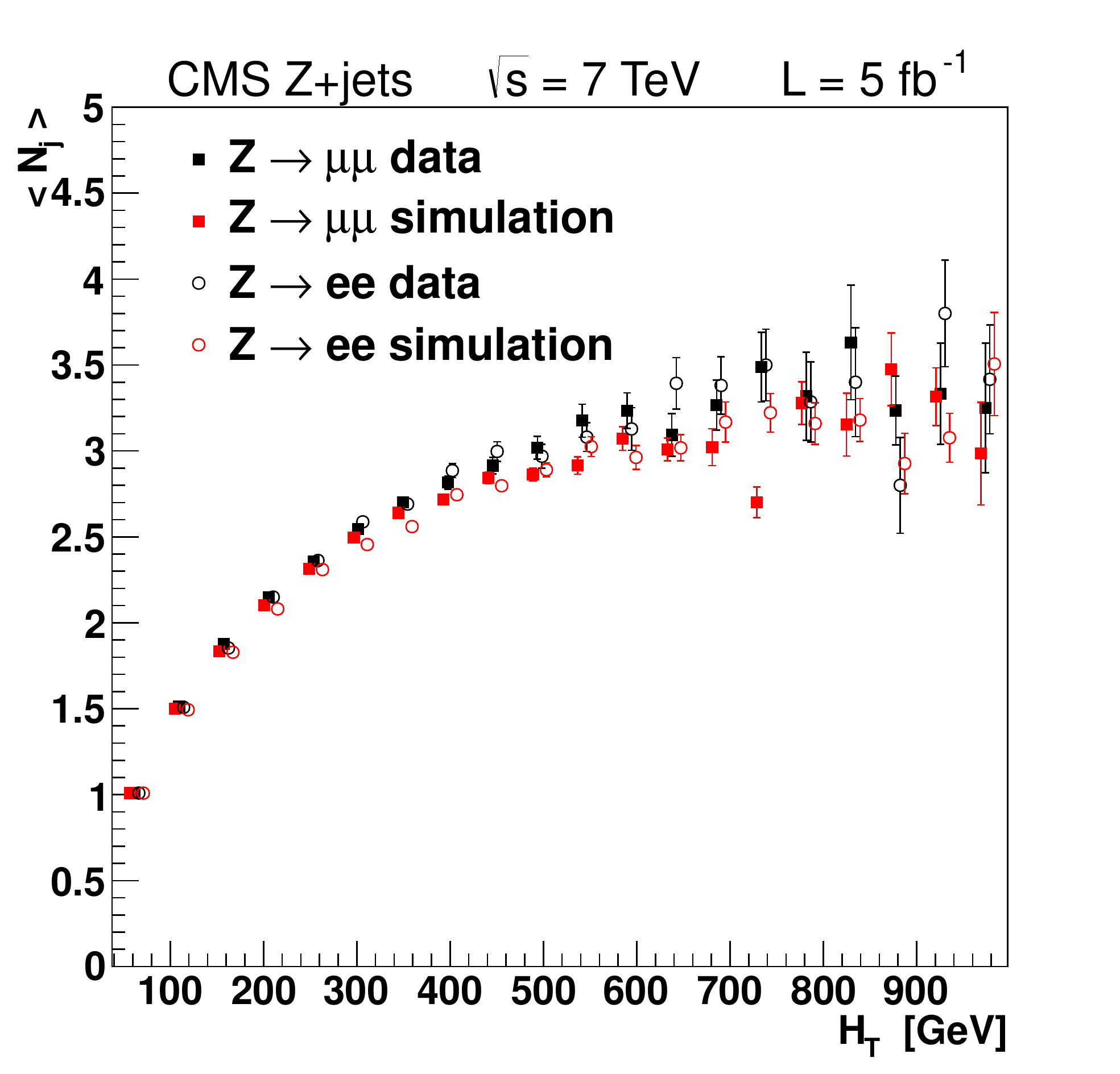}
    \includegraphics[width=7.5cm]{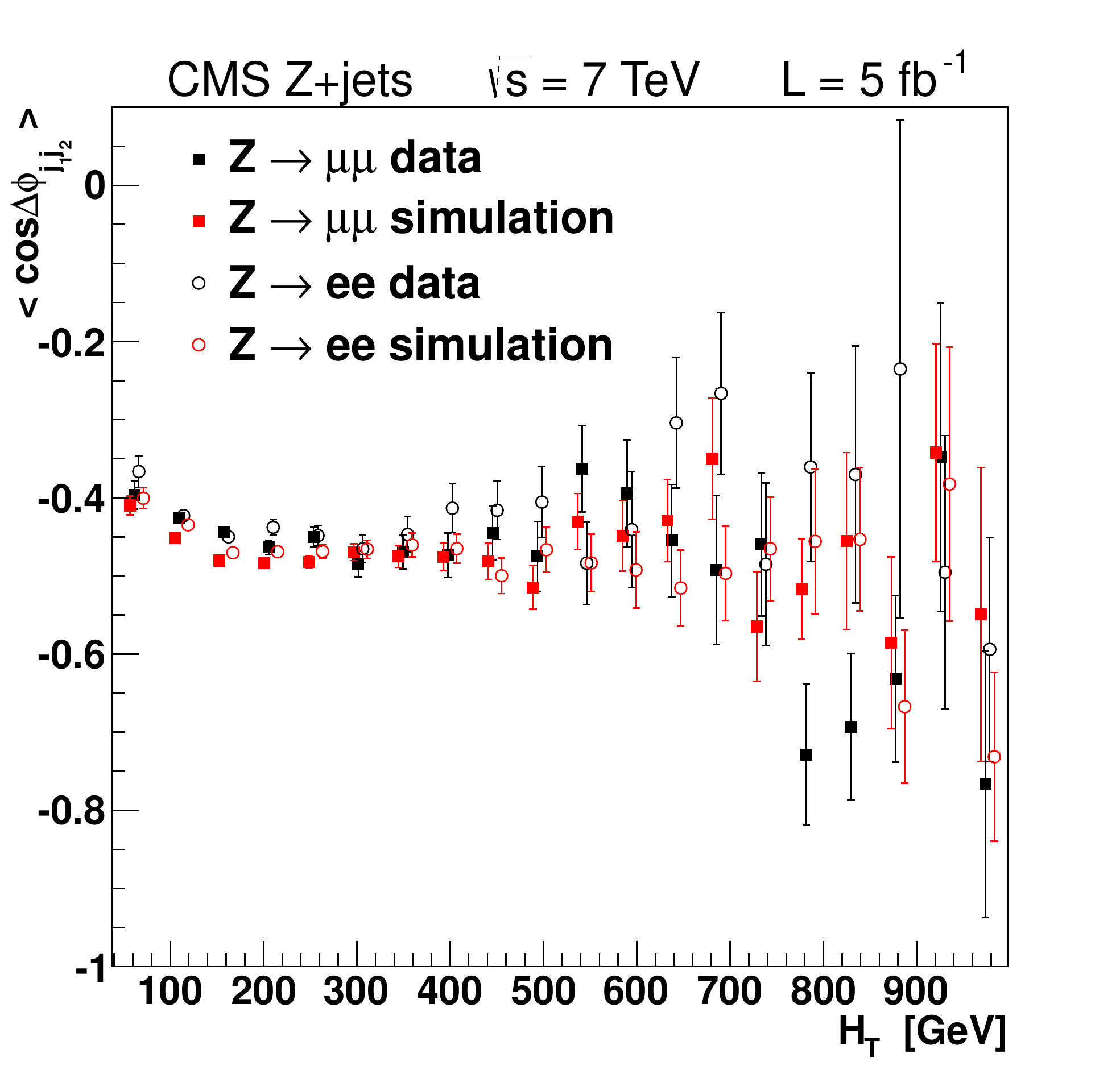}
    \caption{Average number of jets with $\pt > 40\GeV$ as a function of the their total $\HT$
             in Z plus at least one jet events (left) and
             average $\cos\Delta\phi_{\pj_1 \pj_2}$ as a function of the total $\HT$
             in DY Zjj events (right).
             The data points and the points from simulation are shown with
             the statistical uncertainties.}
    \label{fig:Ht_vs}
  \end{center}
\end{figure}

\begin{figure}[hbtp]
  \begin{center}
    \includegraphics[width=7.5cm]{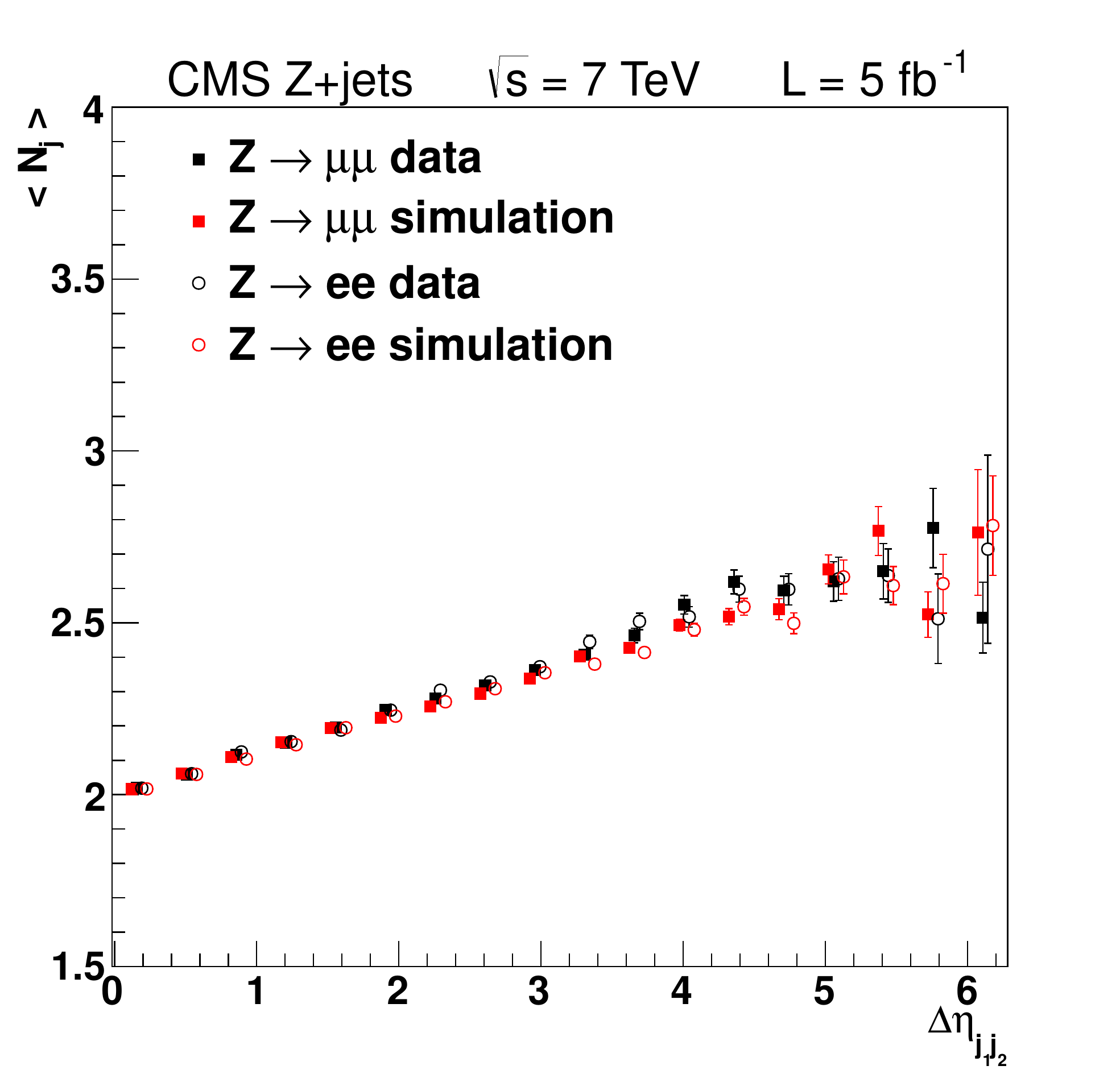}
    \includegraphics[width=7.5cm]{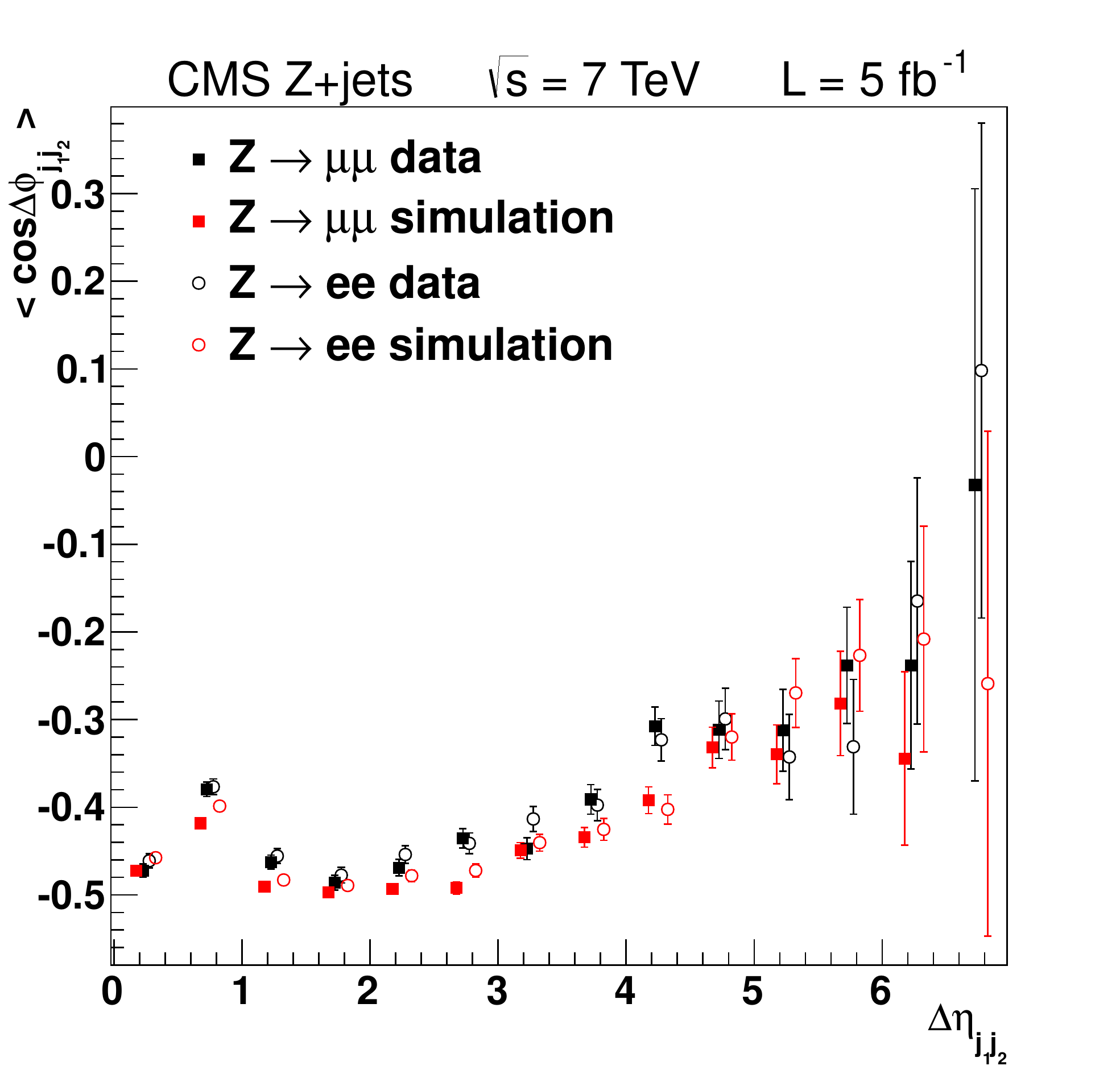}
    \caption{Average number of jets with $\pt > 40\GeV$ as a function of
             $\Delta\eta_{\pj_{1}\pj_{2}}$ (left) and
             average $\cos\Delta\phi_{\pj_{1}\pj_{2}}$ as a function of
             $\Delta\eta_{\pj_{1}\pj_{2}}$ separation (right) in DY Zjj events.
             The data points and the points from simulation are shown with
             the statistical uncertainties.}
    \label{fig:deltaeta_vs}
  \end{center}
\end{figure}

The observables $N_{j}$, $\HT$, $\Delta\eta_{\pj_{1}\pj_{2}}$,
and $\cos \Delta\phi_{\pj_1 \pj_2}$ are measured for jets with $\pt > 40\GeV$.
The events are required to satisfy the $\cPZ_{\Pgm\Pgm}$ and $\cPZ_{\Pe\Pe}$ selection criteria.
Figures~\ref{fig:Ht_vs} and~\ref{fig:deltaeta_vs}
show  the average number of jets and the average $\cos\Delta\phi_{\pj_1 \pj_2}$
as a function of the total $\HT$ and $\Delta\eta_{\pj_{1}\pj_{2}}$. The \MADGRAPH + \PYTHIA (ME-PS)
predictions are in reasonable agreement with the data.

\section{Signal cross section measurement \label{sec:xsection}}
\subsection{Signal extraction using the dijet mass fit \label{sec:analysismjj}}
\begin{figure}[htp]
\begin{center}
\begin{tabular}{c}
\includegraphics[width=0.6\textwidth]{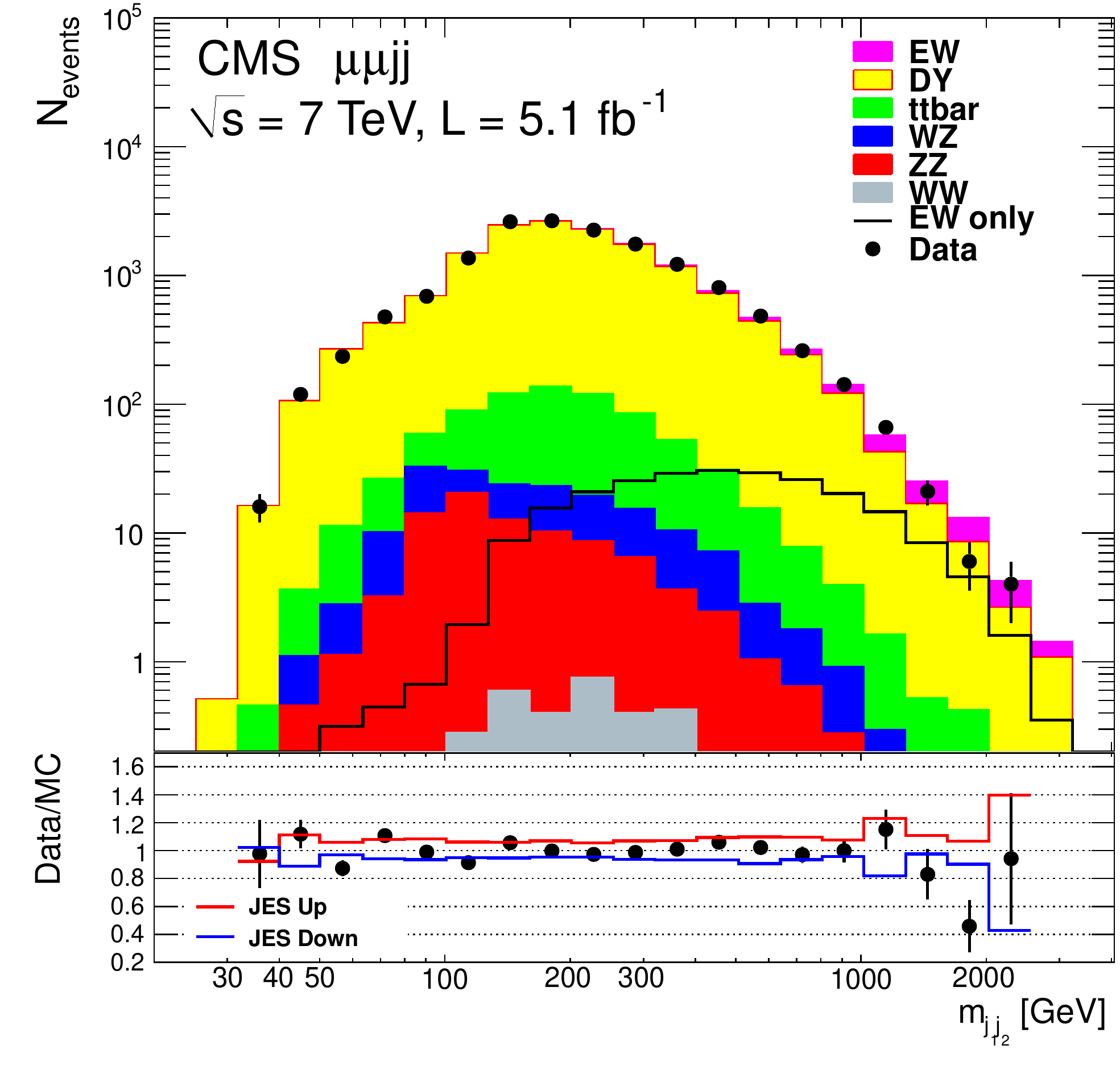}
\end{tabular}
\caption{The $m_{\pj_{1}\pj_{2}}$ distribution after the $\cPZ_{\Pgm\Pgm}$,
         TJ1, and YZ selections (see Section~\ref{sec:selection}).
         The expected contributions from the dominant DY $\Pgm\Pgm \pj\pj $ background and the
         EW $\Pgm\Pgm \pj\pj $ signal processes are evaluated from a fit, while the contributions from
         the small $\ttbar$ and diboson backgrounds are estimated from simulation.
	 The solid line with the label ``EW only'' shows the $m_{\pj_{1}\pj_{2}}$ distribution
         for the signal alone.
         The bottom panel shows the ratio of data over the expected contribution of the signal plus
         background.
	 The region between two lines, with the labels JES Up and JES Down, shows the 1 $\sigma$
         uncertainty due to the jet energy scale uncertainty. The data points are shown
         with the statistical uncertainties.}
\label{fig:mjj_fit}
\end{center}
\end{figure}

The signal cross section in the $\Pgmp\Pgmm$ channel is extracted from a fit of the
$m_{\pj_{1}\pj_{2}}$ data distribution obtained after the $\cPZ_{\Pgm\Pgm}$ selection and
requirements TJ1 and YZ described in Section~\ref{sec:selection}.
The distribution is fitted to the DY $\Pgm\Pgm \pj\pj $
background and the EW $\Pgm\Pgm \pj\pj $ signal processes with MC templates.
Figure~\ref{fig:mjj_fit} shows the $m_{\pj_{1}\pj_{2}}$
distribution where the expected contributions from the dominant DY $\Pgm\Pgm \pj\pj $ background and
the EW $\Pgm\Pgm \pj\pj $ signal are evaluated
from the fit, while the contributions from the small $\ttbar$ and diboson backgrounds are
estimated from simulation.

A likelihood fit with Poisson statistics is performed following the procedure~\cite{fitter}
using the TFractionFitter method in \textsc{root}~\cite{Brun:1997pa}.
The free parameters of the fit, $s$ and $b$, are the ratios of the measured to the expected
event yields of the EW $\Pgm\Pgm \pj\pj $ signal and the DY $\Pgm\Pgm \pj\pj $ background.
The number of expected events is computed in the kinematical region defined 
in Section~\ref{sec:cmsdetrecsim}. 
The numbers of the $\ttbar$  and diboson background events expected from simulation
are fixed in the fit. The fit yields
$s=1.14 \pm 0.28\stat$, $b=0.869 \pm 0.008\stat$ for JPT jets, and
$s=1.14 \pm 0.30\stat$, $b=0.897 \pm 0.008\stat$ for PF jets.
The systematic uncertainties of $s$ are discussed in Section~\ref{sec:systematicsjpt}.

\subsection{Systematic uncertainties \label{sec:systematicsjpt}}

The sources and the absolute values of the systematic uncertainties on the estimated signal value
of $s$ are described below and summarized in Table~\ref{tab:uncertainty_mjj}.

\begin{table}[htp]
\centering
\topcaption{Sources and absolute values of the systematic uncertainties on the estimated
ratio $s$ of measured over expected EW Zjj yields. The simulation of the signal includes
$m_{\pj\pj}> 120\GeV$.}
\begin{tabular}{|l|c|}
\hline
   Source of uncertainty   & Uncertainty \\\hline
  \hline
  \multicolumn{2}{|c|}{Theoretical uncertainties} \\
  \hline
    Background modeling           &  0.20  \\
    Signal modeling               &  0.05  \\
    $\ttbar$ cross section &  0.02   \\
    Diboson cross sections        &  0.01   \\
  \hline
    Total                         &  0.21   \\\hline
  \hline
  \multicolumn{2}{|c|}{Experimental uncertainties} \\
  \hline
    JES+JER                       &  0.44  \\
    Pileup modeling               &  0.05   \\
    MC statistics                 &  0.14  \\
    Dimuon selection              &  0.02   \\
  \hline
    Total                         &  0.47    \\\hline
  \hline
    Luminosity                    &  0.02   \\
\hline
\end{tabular}

\label{tab:uncertainty_mjj}
\end{table}

The following effects are taken into account in the extraction of the signal cross section
from the fit of the $m_{\pj_{1}\pj_{2}}$ distribution:
\begin{itemize}
\item{The theoretical uncertainty on the $m_{\pj_{1}\pj_{2}}$ shape for the dominant
      DY $\Pgm\Pgm \pj\pj $ background process.
      The $m_{\pj_{1}\pj_{2}}$ shape given by the NLO calculation of \MCFM is used
      to correct the shape of \MADGRAPH with jets built from partons and propagated to the
      reconstructed dijet mass with a procedure that matches the reconstructed and the parton jets.
      The fit is then repeated with the modified shape. The systematic uncertainty
      is taken as $s_{\mathrm{NLO}} - s_{\MADGRAPH}$, where
      $s_\mathrm{NLO}$ and $s_{\MADGRAPH}$ are the values of the parameter $s$ extracted from
      the fit
      of the $m_{\pj_{1}\pj_{2}}$ distribution given by \MADGRAPH with and without corrections to the
      NLO shape. The uncertainty of the $m_{\pj_{1}\pj_{2}}$ shape at NLO due to
      the uncertainties in the QCD factorization and renormalization scales, $\mu_\mathrm{F}$ and
      $\mu_\mathrm{R}$, is much smaller than the difference between the shapes given by \MADGRAPH
      and the NLO calculations. The QCD scale in the NLO calculations is varied from $\mu_{0}/2$ to
      $2\mu_{0}$. The $m_{\pj_{1}\pj_{2}}$ shape uncertainty due to the PDFs is found
      to be negligible.}
\item{The theoretical uncertainty of the signal acceptance.
      The acceptance is obtained using the NLO calculation
     \textsc{vbfnlo} as well as using \MADGRAPH. Since
     \textsc{vbfnlo} does not generate events that can be passed through
     the detector simulation, the following parton-level requirements,
     similar to those used in the analysis were applied:
     $\pt^{\ell} > 20\GeV$, $\abs{\eta^{\ell}} < 2.4$, $\pt^\pj  > 50\GeV$, $\abs{\eta^\pj} < 3.6$.
     The acceptance is calculated as the ratio of the cross section with
     parton-level selection to the cross section with the selection
     in the \MADGRAPH simulation of the signal ($m_{\pj\pj}>120\GeV$;
     see Section~\ref{sec:cmsdetrecsim}). The 5\% difference between the
     \textsc{vbfnlo} and \MADGRAPH acceptances is taken as the systematic
     uncertainty. The $m_{\pj\pj}$ shapes given by
     the \textsc{vbfnlo} program and \MADGRAPH simulation are found to be
     very similar, and therefore the shape difference is not included in the
     signal modeling uncertainty. 
     The signal acceptance used in the analysis is evaluated, however, 
     with \MADGRAPH, applying the selections as described in Section~\ref{sec:cmsdetrecsim}.}
\item{The uncertainty on the jet energy scale (JES).
      The $m_{\pj_{1}\pj_{2}}$ fit is repeated with events simulated
      with the jet energy varied by the
      JES uncertainty~\cite{jme10}. The difference between the values of the parameter $s$
      extracted from the fit with simulated events with the adjusted jet
      energy is taken as the systematic uncertainty.}
\item{The uncertainty on the jet energy resolution (JER).
      The $m_{\pj_{1}\pj_{2}}$ fit is repeated with events
      simulated with the correction factor varied by the JER uncertainty~\cite{jme10}.
      The difference between the values of the parameter $s$
      extracted from the fit using simulated events with the adjusted
      data-to-simulation correction factor is taken as the systematic uncertainty.}
\item{The uncertainty on the pileup modeling via re-weighting of the simulated events according
      to the distribution of the number of interactions per beam crossing. The distribution
      is re-evaluated with the total inelastic cross section varied by $\pm$5\%  around the nominal
      value of 68\unit{mb}, based on a set of models consistent with the cross
      section measured by the CMS experiment~\cite{Chatrchyan:2012nj}.}
\item{The uncertainty due to the limited number of events available in the simulated samples
      (MC statistics).}
\item{The uncertainties on the expected yields of $\ttbar$ and diboson events corresponding to
      the theoretical cross section prediction uncertainties.}
\end{itemize}

In addition, the following systematic uncertainties are included in the
estimation of the cross section:
\begin{itemize}
\item{the estimated 2.2\% uncertainty on the integrated luminosity~\cite{smp-12-008},}
\item{the 1\% uncertainty on the data-to-simulation correction factor for the efficiency of
      the lepton reconstruction, identification, isolation, and trigger, which is
      measured with $\cPZ\to \ell \ell$ events.}
\end{itemize}

\subsection{Signal extraction using MVA analysis \label{sec:analysismva}}
The signal is extracted with multivariate analyses in the
$\Pgmp\Pgmm$ and $\Pep\Pem$ channels.
The events are required to pass the $\cPZ_{\Pgm\Pgm}$ or $\cPZ_{\Pe\Pe}$
selection criteria and the tagging jet requirement TJ1.
A boosted decision tree with decorrelation (BDTD option in the
\textsc{tmva} package~\cite{Hocker:2007ht}) is trained to give a high output
value for signal-like events based on the following observables

\begin{itemize}
\item{$\pt^{\pj_{1}}$, $\pt^{\pj_{2}}$, $m_{\pj_{1}\pj_{2}}$,
      $\Delta\eta_{\pj_{1}\pj_{2}}$, and $y^{\ast}$ variables as
      defined in Section~\ref{sec:selection};}
\item{$\pt^{\ell\ell}$: the $\pt$ of the dilepton system;}
\item{$\mathrm{y}_{\ell\ell}$: the rapidity of the dilepton system;}
\item{$\eta_{\pj_{1}}+\eta_{\pj_{2}}$: the sum of the pseudorapidities of the two
      tagging jets;}
\item{$\Delta\phi_{\pj_{1}\pj_{2}}$: the azimuthal separation of the
      two tagging jets;}
\item{$\Delta\phi({\ell\ell},\pj_{1})$ and $\Delta\phi({\ell\ell},\pj_{2})$:
      the azimuthal separations between the dilepton system and the two
      tagging jets.}
\end{itemize}

In the $\Pep\Pem$ channel the gluon-quark likelihood values for the tagging
jets are also used as inputs. In the DY $\ell\ell \pj\pj $ background about 50\% of the jets
originate from gluons while in the EW $\ell\ell \pj\pj $ signal process the tagging jets are
only initiated by quarks. A likelihood discriminator separates the gluon-originated jets from the
quark-originated jets.
The discriminator makes use of five internal jet properties, built from the
jet constituents. These are related to the two angular spreads (root mean square)
of the constituents in the $\eta$-$\phi$ plane, the asymmetry (pull) of the constituents
with respect to the center of the jet,
the multiplicity of the constituents, and the maximum energy fraction carried by a single
constituent. The validations of the five input variables and of the gluon-quark
likelihood output have been carried out using the multijet, Z+jet, and photon+jet
samples, for which the relative differences between data and simulation are within 10\%.
To assess the systematic uncertainty from the usage of this tool, the gluon likelihood
output in the simulated samples has been modified in accord with the differences observed
in the three samples. The use of the gluon-quark likelihood discriminator leads to a
decrease of the statistical uncertainty of the measured signal in the $\Pep\Pem$ channel
by 5\%.

\begin{figure}[htp]
\begin{center}
\includegraphics[width=0.49\textwidth]{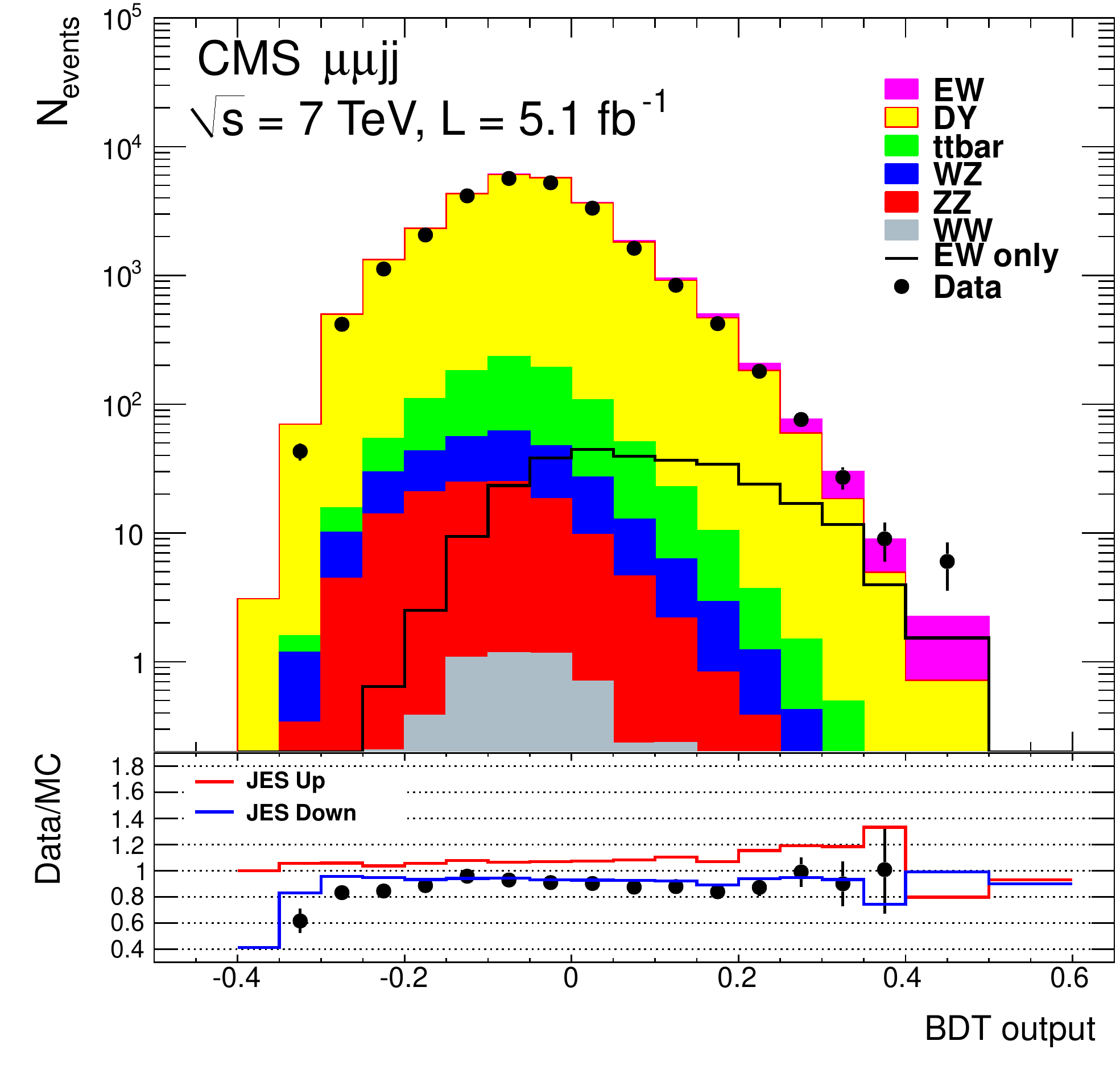}
\includegraphics[width=0.49\textwidth]{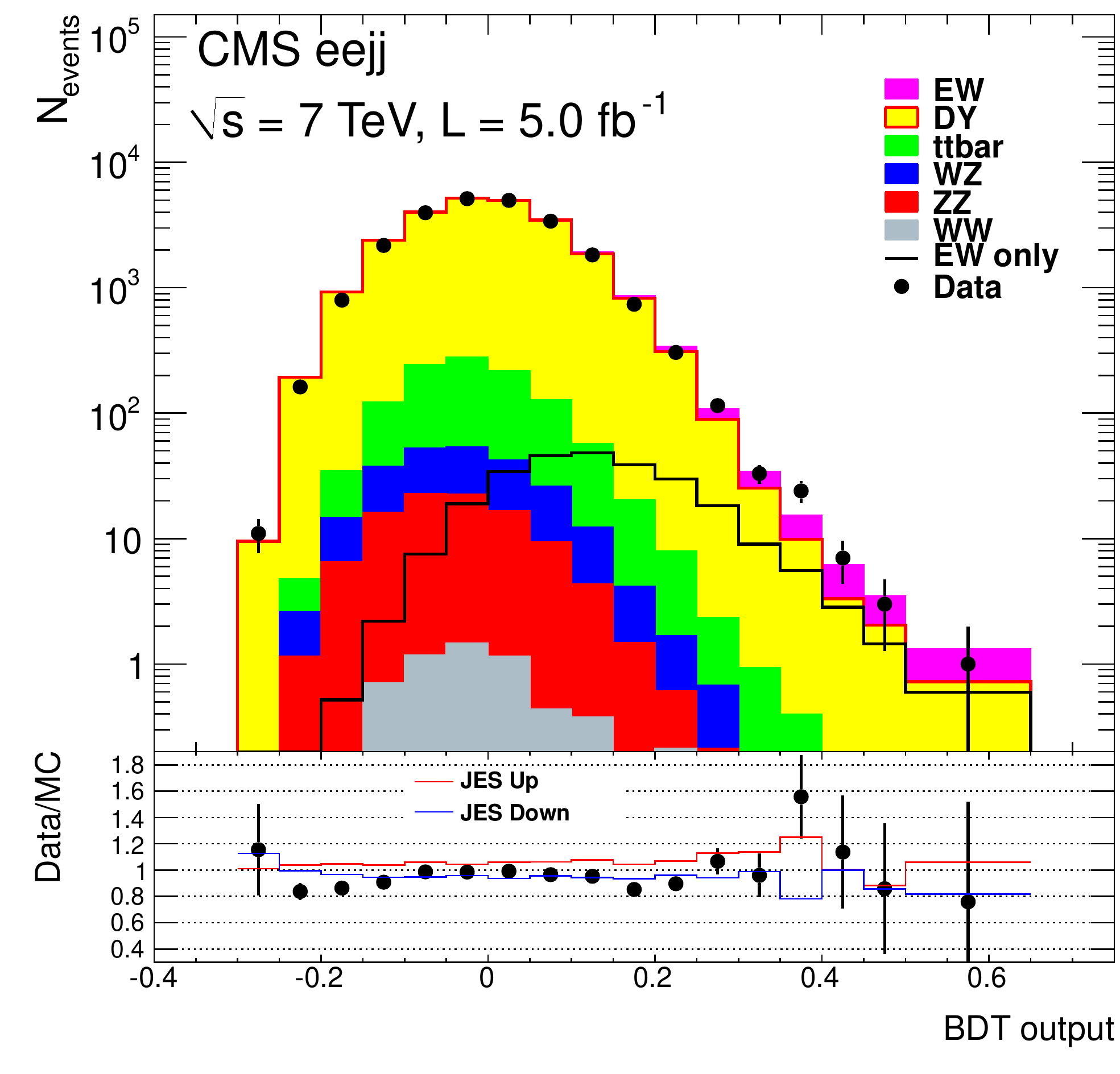}
\caption{The BDT output distributions for the $\Pgmp\Pgmm$ channel (left) obtained with
         JPT jets and for the $\Pep\Pem$ channel (right) obtained with PF jets
         after applying the $\cPZ_{\Pgm\Pgm}$ and $\cPZ_{\Pe\Pe}$ selection
         criteria, respectively, with the tagging jet requirement TJ1. The expected
         contributions from the signal and background processes are evaluated from
         simulation.
	 The solid line with the label ``EW only'' shows the BDT output distribution
         for the signal alone.
         The bottom panels show the ratio of data over the expected
         contribution of the signal plus background.
	 The region between the two lines, with the labels JES Up and JES Down,
         shows the $1\,\sigma$
         uncertainty of the simulation prediction due to the jet energy scale uncertainty.
         The data points are shown with the statistical uncertainties.}
\label{fig:bdt_nfit}
\end{center}
\end{figure}

The BDT is trained with EW $\ell\ell \pj\pj $ simulated events for the signal model along with the
DY $\ell\ell \pj\pj $ and $\ttbar$ simulated events for the background model.
The BDT output value is proportional to the probability that the event belongs to
the signal: the higher the value, the higher the probability. The BDT output distributions for the two
lepton modes from various production mechanisms are shown in Fig.~\ref{fig:bdt_nfit} where the
expected contributions from the signal and background processes are evaluated from simulation.

The signal cross section is extracted from the fit of the BDT
output distributions for data with the method described in Section~\ref{sec:analysismjj},
for the $m_{\pj_{1}\pj_{2}}$ distributions.
For the $\Pgmp\Pgmm$ channel,
the best fits are
$s=0.90 \pm 0.19\stat$, $b=0.905 \pm 0.006\stat$ with JPT jets and
$s=0.85 \pm 0.18\stat$, $b=0.937 \pm 0.007\stat$ with PF jets.
For the $\Pep\Pem$  channel, with PF jets, the best fit is
$s=1.17 \pm 0.27\stat$, $b=0.957 \pm 0.010\stat$.
The value of the parameter $b$ obtained from the fit is below unity by 5--10\%. It is however
consistent with unity within the JES uncertainty and the systematic uncertainty in
the \MADGRAPH simulation of the DY $\ell\ell \pj\pj $ process as discussed in
Section~\ref{sec:selection}.

\begin{figure}[htp]
\begin{center}
\includegraphics[width=0.49\textwidth]{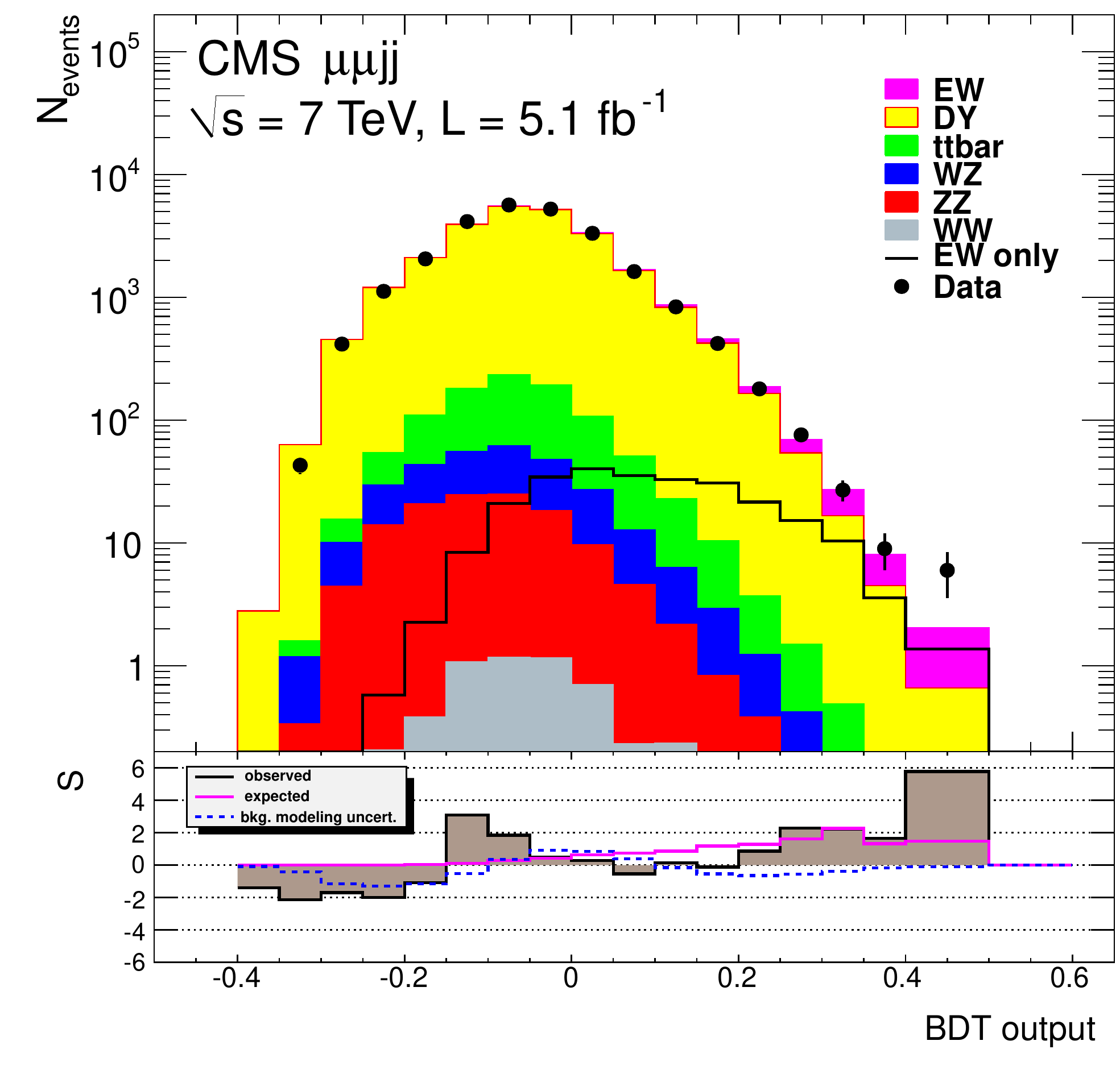}
\includegraphics[width=0.49\textwidth]{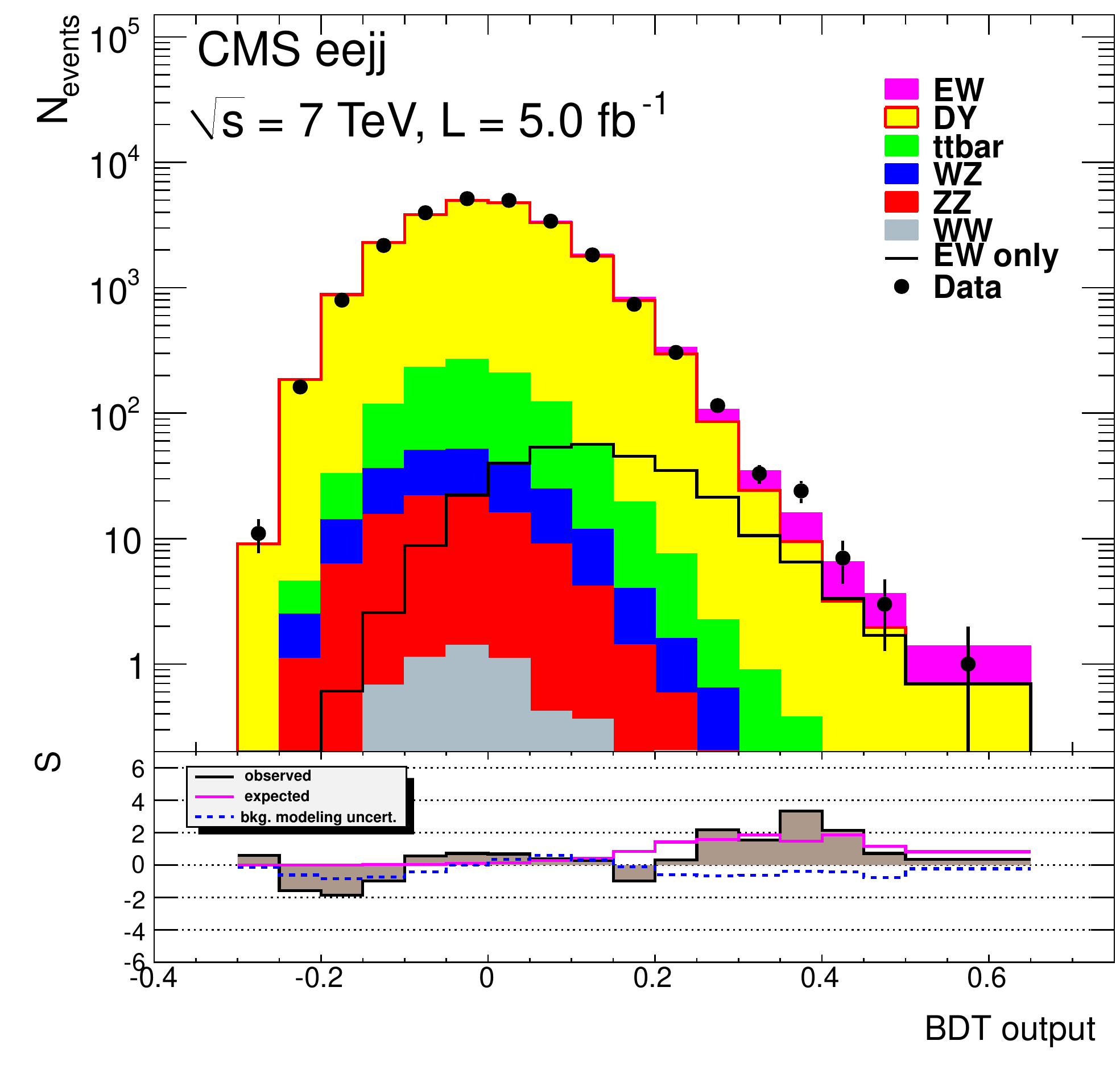}
\caption{The BDT output distributions for the $\Pgmp\Pgmm$ channel (left) obtained with
         JPT jets and for the $\Pep\Pem$ channel (right) obtained with PF jets
         after the respective $\cPZ_{\Pgm\Pgm}$ and $\cPZ_{\Pe\Pe}$ selections
         and the tagging jet requirement TJ1.
         The expected contributions from the dominant DY $\ell\ell \pj\pj $
         background and the EW $\ell\ell \pj\pj $ signal processes are evaluated
         from the fit. The contributions from the small $\ttbar$ and diboson backgrounds
         are estimated from simulation.
	 The solid line with the label ``EW only'' shows the BDT output distribution
         for the signal alone.
         The bottom panels show the significance
         observed in data (histogram) and expected from simulation (solid purple line).
         The dashed blue line shows the background modeling uncertainty. The calculation
         of the significance and background modeling uncertainty are explained in the text.
         The data points are shown with the statistical uncertainties.}
\label{fig:bdt_fit}
\end{center}
\end{figure}

Figure~\ref{fig:bdt_fit} shows the BDT output distributions for the $\Pgmp\Pgmm$ (left)
and $\Pep\Pem$ (right) channels, where the expected contributions from the dominant
DY $\ell\ell \pj\pj $ background and the EW $\ell\ell \pj\pj $ signal processes are evaluated
from the fit; the contributions from the small $\ttbar$ and diboson backgrounds are
taken from the simulation estimates.

The presence of the signal is clearly seen at high
values of the BDT output ($>$0.25) for both dimuon and dielectron channels, and in the cases
when the dominant DY $\ell\ell \pj\pj $ background is evaluated from simulation
(Fig.~\ref{fig:bdt_nfit}) or from the fit (Fig.~\ref{fig:bdt_fit}).

In Fig.~\ref{fig:bdt_fit} the bottom panels show the
significance observed in data (histogram) and expected from simulation (solid purple line),
while the dashed blue line shows the background modeling uncertainty. The observed
signal significance in bin $i$ of the BDT output distribution is calculated as

\begin{equation}
  S^{\text{observed}}_{i}=
  \frac{N^{\text{data}}_{{i}}-N^{\text{bkg}}_{i}}
  {\sqrt{N^{\text{bkg}}_{{i}}+\left(\Delta B_{{i}}^{\mathrm{JES}}\right)^{2}}},
\end{equation}

where $N^{\text{data}}_{i}$ and $N^{\text{bkg}}_{{i}}$ are the
number of the observed events and the number of the simulated background events obtained from
the fit, respectively. $\Delta B_{{i}}^{\mathrm{JES}}$ is the dominant experimental
systematic uncertainty due to the JES, calculated as

\begin{equation}
  \Delta B_{i}^{\mathrm{JES}} =
  \sqrt{0.5
    \left[\left(N_{{i}}^{\text{bkg}}-N_{i,\mathrm{JESup}}^{\text{bkg}}\right)^{2}+
     \left(N_{{i}}^{\text{bkg}}-N_{i,\mathrm{JESdn}}^{\text{bkg}}\right)^{2}\right]},
\end{equation}

where $N_{i,\mathrm{JESup}}^{\text{bkg}}$ and $N_{i,\mathrm{JESdn}}^{\text{bkg}}$
are the numbers of the simulated background events from the fit with the jet
energy varied by the JES uncertainty.
The expected signal significance is calculated as

\begin{equation}
  S^{\text{expected}}_{{i}}=
  \frac{N^{\mathrm{EW~Zjj}}_{{i}}}
  {\sqrt{N^{\text{bkg}}_{{i}}+\left(\Delta B_{{i}}^{\mathrm{JES}}\right)^{2}}},
\end{equation}

where $N^{\mathrm{EW~Zjj}}_{{i}}$ is the number of simulated signal EW Zjj events
from the fit.

The background modeling uncertainty is calculated as

\begin{equation}
  \left(N^{{\MCFM}}_{{i}}-N^{\text{bkg}}_{{i}}\right)\Big/
  \sqrt{N^{\text{bkg}}_{{i}}+\left(\Delta B_{{i}}^{\mathrm{JES}}\right)^{2}},
\end{equation}

where $N^{{\MCFM}}_{i}$ is the number of the simulated background events obtained
from a new fit. The fit uses a modified BDT output distribution for the DY $\ell\ell \pj\pj $
process. This distribution is evaluated using the $m_{\pj_{1} \pj_{2}}$ shape
obtained from the NLO calculation of \MCFM, as explained in Section~\ref{sec:systematicsjpt}.

The sources of the systematic uncertainties on the estimated signal value of $s$ are
those discussed in Section~\ref{sec:systematicsjpt}. The absolute values of the systematic
uncertainties on the value of $s$ for the BDT analysis are shown in
Table~\ref{tab:uncertainty_mva} for the $\Pgmp\Pgmm$ and $\Pep\Pem$ modes.
The uncertainties are smaller than those from the $m_{\pj_{1}\pj_{2}}$ fit analysis since the BDT
approach provides better separation between signal and background.

\begin{table}[htp]
\centering
\caption{The sources and absolute values of systematic uncertainties on the estimated
        ratio $s$ of measured over expected EW Zjj yields
 in the BDT analysis for the $\Pgmp\Pgmm$ and $\Pep\Pem$
         channels.}
\begin{tabular}{|l|c|c|}
\hline
   Source of uncertainty   & \multicolumn{2}{c|}{Uncertainty} \\\cline{2-3}
                          &  $\Pgmp\Pgmm$ channel  &  $\Pep\Pem$ channel \\\hline
  \hline
  \multicolumn{3}{|c|}{Theoretical uncertainties} \\
  \hline
    Background modeling           &  0.15           &  0.16    \\
    Signal modeling               &  0.05           &  0.05    \\
    $\ttbar$ cross section &  0.03           &  0.03     \\
    Diboson cross sections        &  0.02           &  0.02     \\
  \hline
   Total                          &  0.16           &  0.17    \\\hline
  \hline
  \multicolumn{3}{|c|}{Experimental uncertainties} \\
  \hline
    JES+JER                       &  0.22           &  0.29    \\
    Pileup modeling               &  0.03           &  0.03    \\
    MC statistics                 &  0.13           &  0.19    \\
    Gluon-quark discriminator     &  not used       &  0.02    \\
    Dilepton selection            &  0.02           &  0.02     \\
  \hline
    Total                         &  0.26           &  0.35     \\\hline
  \hline
    Luminosity                    &  0.02           &  0.03     \\
\hline
\end{tabular}

\label{tab:uncertainty_mva}
\end{table}

The BDT analysis in the $\Pgmp\Pgmm$ channel is repeated for events
passing the additional requirement of $\abs{{y}^{\ast}} < 1.2$,
as used in the $m_{\pj_{1}\pj_{2}}$ analysis. In this case the best fit values are:
$s=1.50 \pm 0.26\stat$, $b=0.863 \pm 0.007\stat$ for the analysis with JPT jets and
$s=1.37 \pm 0.25\stat$, $b=0.862 \pm 0.007\stat$ for the analysis
with PF jets. These values are compatible
with those obtained from the method based on the $m_{\pj_{1}\pj_{2}}$ fit, as described in
Section~\ref{sec:analysismjj}.

\subsection{Results \label{sec:resultjpt}}

The presence of the signal is confirmed in the dimuon and dielectron channels
by using two alternative jet reconstruction algorithms and two methods of signal extraction.

The BDT analysis provides smaller uncertainties on the parameter $s$, and therefore
the result is based on this analysis.
The measured cross section is
$\sigma_\text{meas}=s \times \sigma_{\MADGRAPH}(\mathrm{EW}~\ell\ell \pj\pj )$, where
$\sigma_{\MADGRAPH}(\mathrm{EW}~\ell\ell \pj\pj )=162\unit{fb}$ per lepton
flavor is the cross section obtained from the \MADGRAPH simulation using CTEQ6L1~\cite{Pumplin:2002vw}.

The signal cross section given by \MADGRAPH is obtained for event generation
with the following selections at the parton level:
$m_{\ell\ell}>50\GeV$, $\pt^\pj  > 25\GeV$, $\abs{\eta ^\pj} < 4.0$, $m_{\pj\pj} > 120\GeV$.
The parton--level requirements on the jet $\pt$ and $\eta$ maximize the signal selection
efficiency relative to the actual selection applied to the data, while keeping the fraction of
the events which fail the parton--level requirements but pass the data selection criteria
at a negligible level.

The cross section for the dimuon mode with JPT jets is:

\begin{equation}
\label{eq:xsect_mumu_jpt}
\sigma_{\Pgm\Pgm}^{\mathrm{EW}}~(\mathrm{JPT}) = 146
                            \pm 31\stat
                            \pm 42\,\text{(exp. syst.)}
                            \pm 26\,\text{(th. syst.)}
		            \pm 3\lum\unit{fb}.
\end{equation}

The cross section for the dimuon mode with PF jets is

\begin{equation}
\label{eq:xsect_mumu_pf}
\sigma_{\Pgm\Pgm}^{\mathrm{EW}}~(\mathrm{PF}) = 138
                            \pm 29\stat
                            \pm 40\,\text{(exp. syst.)}
                            \pm 25\,\text{(th. syst.)}
		            \pm 3\lum\unit{fb}.
\end{equation}

The measurements for the dimuon mode, with the two different jet reconstruction algorithms, are
compatible.

In the dielectron mode with PF jets, the cross section is:

\begin{equation}
\label{eq:xsect_ee}
\sigma^{\mathrm{EW}}_{\Pe\Pe}\,\mathrm{(PF)} = 190
                       \pm 44\stat
                       \pm 57\,\text{(exp. syst.)}
                       \pm 27\,\text{(th. syst.)}
		       \pm 4\lum\unit{fb}.
\end{equation}

The measured cross sections agree with the theoretical value of
$\sigma_\textsc{vbfnlo}(\mathrm{EW }\;\ell\ell\pj\pj ) =166$\unit{fb},
calculated with next-to-leading order QCD corrections using the same parton level
selections as those applied in the signal event generation by \MADGRAPH.
The cross sections obtained in the $\Pgmp\Pgmm$ and $\Pep\Pem$
analyses using PF jets is combined and
the average cross section
is:

\begin{equation}
\label{eq:xsect_comb}
\sigma^{\mathrm{EW}}_{\ell\ell~(\ell=\Pe,\,\mu)} = 154
                       \pm 24\stat
                       \pm 46\,\text{(exp. syst.)}
                       \pm 27\,\text{(th. syst.)}
		       \pm 3\lum\unit{fb} .
\end{equation}

\section{Summary \label{sec:conclusions}}
A measurement of the electroweak production of a Z boson in association with two jets
in pp collisions at $\sqrt{s} = 7\TeV$ has been carried out with the CMS detector
using an integrated luminosity of 5\fbinv.
The cross section for the
EW $\ell\ell \pj\pj $ ($\ell = \Pe$,\,$\mu$) production process, with
$m_{\ell\ell} >50\GeV$, $\pt^\pj  > 25\GeV$, $\abs{\eta ^\pj }< 4.0$,
$m_{\pj\pj} >120\GeV$, is
                $\sigma = 154
                \pm 24\stat
                \pm 46\,\text{(exp. syst.)}
                \pm 27\,\text{(th. syst.)}
                \pm 3\lum\unit{fb}.$
The measurement is in agreement with the theoretical cross section of 166\unit{fb},
obtained with calculations including next-to-leading order QCD corrections based on the
CT10~\cite{Lai:2010vv} parton distribution functions.
A significance of 2.6 standard deviations has been obtained for
the observation of EW production of the Z boson with two tagging jets.
The measured hadronic activity in events with Drell--Yan production in association with two
jets is in good agreement with simulation. This is the first measurement of EW production of a
Z boson with two jets at a hadron collider, and constitutes an important foundation for the more
general study of vector boson fusion processes, of relevance for Higgs boson searches and for
measurements of electroweak gauge couplings and vector-boson scattering.

\section*{Acknowledgments \label{sec:acknowledgements}}
We congratulate our colleagues in the CERN accelerator departments for the excellent performance of the LHC and thank the technical and administrative staffs at CERN and at other CMS institutes for their contributions to the success of the CMS effort. In addition, we gratefully acknowledge the computing centers and personnel of the Worldwide LHC Computing Grid for delivering so effectively the computing infrastructure essential to our analyses. Finally, we acknowledge the enduring support for the construction and operation of the LHC and the CMS detector provided by the following funding agencies: BMWF and FWF (Austria); FNRS and FWO (Belgium); CNPq, CAPES, FAPERJ, and FAPESP (Brazil); MEYS (Bulgaria); CERN; CAS, MoST, and NSFC (China); COLCIENCIAS (Colombia); MSES (Croatia); RPF (Cyprus); MoER, SF0690030s09 and ERDF (Estonia); Academy of Finland, MEC, and HIP (Finland); CEA and CNRS/IN2P3 (France); BMBF, DFG, and HGF (Germany); GSRT (Greece); OTKA and NKTH (Hungary); DAE and DST (India); IPM (Iran); SFI (Ireland); INFN (Italy); NRF and WCU (Republic of Korea); LAS (Lithuania); CINVESTAV, CONACYT, SEP, and UASLP-FAI (Mexico); MSI (New Zealand); PAEC (Pakistan); MSHE and NSC (Poland); FCT (Portugal); JINR (Armenia, Belarus, Georgia, Ukraine, Uzbekistan); MON, RosAtom, RAS and RFBR (Russia); MSTD (Serbia); SEIDI and CPAN (Spain); Swiss Funding Agencies (Switzerland); NSC (Taipei); ThEPCenter, IPST and NSTDA (Thailand); TUBITAK and TAEK (Turkey); NASU (Ukraine); STFC (United Kingdom); DOE and NSF (USA).

Individuals have received support from the Marie-Curie programme and the European Research Council and EPLANET (European Union); the Leventis Foundation; the A. P. Sloan Foundation; the Alexander von Humboldt Foundation; the Belgian Federal Science Policy Office; the Fonds pour la Formation \`a la Recherche dans l'Industrie et dans l'Agriculture (FRIA-Belgium); the Agentschap voor Innovatie door Wetenschap en Technologie (IWT-Belgium); the Ministry of Education, Youth and Sports (MEYS) of Czech Republic; the Council of Science and Industrial Research, India; the Compagnia di San Paolo (Torino); the HOMING PLUS programme of Foundation for Polish Science, cofinanced by EU, Regional Development Fund; and the Thalis and Aristeia programmes cofinanced by EU-ESF and the Greek NSRF.
\bibliography{auto_generated}   

\cleardoublepage \appendix\section{The CMS Collaboration \label{app:collab}}\begin{sloppypar}\hyphenpenalty=5000\widowpenalty=500\clubpenalty=5000\textbf{Yerevan Physics Institute,  Yerevan,  Armenia}\\*[0pt]
S.~Chatrchyan, V.~Khachatryan, A.M.~Sirunyan, A.~Tumasyan
\vskip\cmsinstskip
\textbf{Institut f\"{u}r Hochenergiephysik der OeAW,  Wien,  Austria}\\*[0pt]
W.~Adam, T.~Bergauer, M.~Dragicevic, J.~Er\"{o}, C.~Fabjan\cmsAuthorMark{1}, M.~Friedl, R.~Fr\"{u}hwirth\cmsAuthorMark{1}, V.M.~Ghete, N.~H\"{o}rmann, J.~Hrubec, M.~Jeitler\cmsAuthorMark{1}, W.~Kiesenhofer, V.~Kn\"{u}nz, M.~Krammer\cmsAuthorMark{1}, I.~Kr\"{a}tschmer, D.~Liko, I.~Mikulec, D.~Rabady\cmsAuthorMark{2}, B.~Rahbaran, C.~Rohringer, H.~Rohringer, R.~Sch\"{o}fbeck, J.~Strauss, A.~Taurok, W.~Treberer-Treberspurg, W.~Waltenberger, C.-E.~Wulz\cmsAuthorMark{1}
% ...
\vskip\cmsinstskip
% ...
1:~~Also at Vienna University of Technology, Vienna, Austria\\
2:~~Also at CERN, European Organization for Nuclear Research, Geneva, Switzerland\\
% ...

\end{sloppypar}
\end{document}